\documentclass[aps,amsmath,nofootinbib,amssymb,superscriptaddress]{revtex4}
\usepackage{amssymb,graphicx,color,dcolumn,bm,slashed,amsmath,here}
\usepackage[title]{appendix}

\begin{document}
{\hspace{+14cm} KEK-TH-2175}
\title{Probing the top Yukawa coupling at the LHC via \\ associated production of single top and Higgs}
\author{Vernon Barger}\email {barger@pheno.wisc.edu}
\affiliation{Department of Physics, University of Wisconsin, Madison, WI 53706, U.S.A.}
\author{Kaoru Hagiwara} \email{kaoru.hagiwara@kek.jp}
\affiliation{Tsung-Dao Lee Institute, Shanghai Jiao Tong University,
Shanghai 200240, China}
\affiliation{KEK Theory Center and Sokendai, Tsukuba, Ibaraki 305-0801, Japan}
\author{Ya-Juan Zheng} \email{yjzheng@ku.edu}
\affiliation{Department of Physics and Astronomy, University of Kansas, Lawrence, KS 66045, U.S.A.}

\begin{abstract}
We study Higgs boson production associated with single top or anti-top via $t$-channel weak boson exchange at the LHC.
 The process is an ideal probe of the top quark Yukawa coupling, because we can measure the relative phase of  $htt$ and $hWW$ couplings, thanks to the significant interference between the two amplitudes. 
By choosing the emitted $W$ momentum along the polar axis in the $th\,(\bar{t}h)$ rest frame, we obtain the helicity amplitudes for all the contributing subprocesses analytically, with possible CP phase of the Yukawa coupling. 
We study the azimuthal asymmetry between the $W$ emission and the $Wb\,(\bar{b})\to t(\bar{t})\,h$ scattering planes, as well as several $t$ and $\bar{t}$ polarization asymmetries as a signal of CP violating phase in the $htt$ coupling. 
Both the azimuthal asymmetry and the polarization perpendicular to the scattering plane are found to have the opposite sign between the top and anti-top events. 
We identify the origin of the sign of asymmetries, and propose the possibility of direct CP violation test in $pp$ collisions by comparing the top and anti-top polarization at the LHC.
\end{abstract}

\maketitle

\clearpage
\section{Introduction}

The top quark Yukawa coupling of the 125 GeV Higgs boson 
($h$) is the largest of the Standard Model (SM) couplings, 
and the precise measurement of its magnitude and properties 
is the important target of the LHC experiments.   
Measurements of the loop-induced $hgg$ and $h\gamma\gamma$ 
transitions constrain the top Yukawa, or $htt$ coupling 
indirectly, if only the SM particles contribute to the 
vertices with the SM couplings.  
The observation of the associated production of the Higgs 
boson and the top quark pair at the LHC\,\cite{Sirunyan:2018hoz,Aaboud:2018urx} determines the $htt$ coupling directly, constraining 
its magnitude to be within about 10\% of the SM prediction. 

In this paper, we study the possibility of measuring 
a possible CP violating phase of the $htt$ coupling in 
the Higgs boson production associated with single top 
or anti-top at the LHC.  
The cross section is dominated by the so-called $t$-channel 
$W$ exchange process, where the $W$ boson emitted from 
a quark or anti-quark in a proton scatters with a  
$b$ or $\bar{b}$ quark in the other proton to produce 
a pair of $h$ and $t$, or $\bar{t}$.    
The process is particularly sensitive to the phase of 
the $htt$ coupling, because we can measure the real 
and imaginary part of the $htt$ coupling through the 
interference between the amplitudes with the $htt$ 
and $hWW$ couplings which have the same order of magnitude  
with opposite sign\,\cite{Stirling:1992fx,Bordes:1992jy} 
in the SM limit.  
We can therefore measure the phase of the $htt$ coupling 
with respect to that of the $hWW$ coupling, whose magnitude 
and phase have already been constrained rather well\,\cite{Aaboud:2018jqu,Aad:2019lpq,Sirunyan:2018egh}  
and will be determined precisely in the HL-LHC era.  

We adopt the following minimal non-SM modification to the 
top Yukawa coupling,
\begin{eqnarray}          
  \label{eq:coupling} 
{\cal L}_{htt} =
-g_{htt}h\bar{t}(\cos\xi_{htt}+i\sin\xi_{htt}\gamma_5)t
=
-g_{htt} h  \{ e^{-i\xi_{htt}} t_R^\dagger t_L 
            + e^{ i\xi_{htt}} t_L^\dagger t_R \},
\end{eqnarray}
where we introduce the positive $\kappa$ factor as 
\begin{eqnarray}
g_{htt}^{} = (m_t/v) \kappa_{htt} ^{}> 0 
\end{eqnarray}
for the normalization of the coupling.  
The Lagrangian expressed in terms of the chiral two-spinors 
$t_L$ and $t_R$ 
\begin{eqnarray}
\frac{1-\gamma_5}{2} t = 
\left(
{\begin{array}{*{20}{c}}
t_L\\
0
\end{array}}
\right),
\quad\quad
\frac{1+\gamma_5}{2} t = 
\left(
{\begin{array}{*{20}{c}}
0\\
t_R
\end{array}}
\right),
\end{eqnarray}
show that $\xi_{htt}$ is the CP phase of the Yukawa 
interactions.  
Its defined range is 
\begin{eqnarray}
-\pi < \xi_{htt} \lesssim \pi 
\end{eqnarray}
with respect to the $hWW$ coupling term 
\begin{eqnarray}
{\cal L}_{hWW}^{} &=& g_{hWW}^{} h W^-_\mu W^{+\mu} 
\label{eq:lhww}
\end{eqnarray}
for which we take the real positive value 
\begin{eqnarray}
\label{eq:ghww}
g_{hWW}^{} = (2 m_W^2/v) \kappa_{hWW}^{}>0.   
\end{eqnarray}

CP violation in the $htt$ coupling,   
$\xi_{htt} \neq 0$, with $\kappa_{htt} \neq 1$  
can arise by radiative effects in the $htt$ vertex 
due to new interactions which violate CP, 
or in models with two or more Higgs doublets  
when the Higgs interactions violate CP.  
Once the underlying new physics model is fixed, 
we often obtain correlations among the non-SM effective 
couplings, such as $\kappa_{hWW}^{}$, $\kappa_{htt}^{}$, 
$\xi_{htt}^{}$, and also for the other $hff$ couplings 
as well as the loop induced $hgg$, $h\gamma\gamma$ 
and $hZ\gamma$ vertices.  
In this report, we set 
\begin{eqnarray}
\kappa_{htt}^{}=\kappa_{hWW}^{}=1 
\end{eqnarray} 
in all the numerical results, in order to focus 
on the observable CP violating effects for relatively 
small phase 
\begin{eqnarray}
|\xi_{htt}| \lesssim 0.1\pi.  
\end{eqnarray}

\begin{figure}[h]
\begin{centering}
\begin{tabular}{c}
\includegraphics[width=0.45\textwidth]{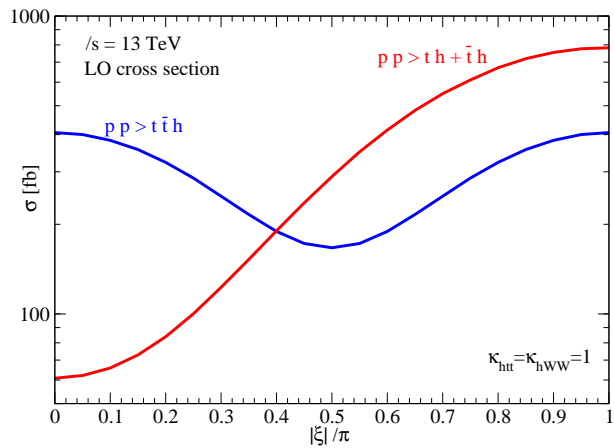}\label{fig:1a}
\end{tabular}
\caption {
LO cross section at the LHC ($\sqrt{s}=13$~TeV) for the sum 
of $pp\to th$ and $pp\to \bar{t}h$ production via 
$t$-channel $W$ exchange as a function of the CP phase 
$|\xi_{htt}|$ for $\kappa_{htt}^{}=1$.  
Also shown is the $pp\to t\bar{t}h$ production cross section 
in the same model. 
}
\label{fig:cs}
\end{centering}
\end{figure}

In Fig.\,\ref{fig:cs}, we show the total cross section of the Higgs 
boson production with single $t$ or $\bar{t}$ via $t$-channel 
$W$ exchange in $pp$ collisions at $\sqrt{s}=13$~TeV 
for the effective $htt$ coupling of Eq.\,(\ref{eq:coupling}), with 
$\kappa_{htt}=1$ and $|\xi_{htt}|$ between $0$ (SM) and $\pi$.
Also shown is the total cross section for $h$ and a 
$t\bar{t}$ pair in the same model.
They are obtained by MadGraph\,\cite{Alwall:2014hca} 
with the effective Lagrangian of Eq.\,(\ref{eq:coupling}) in 
Feynrules\,\cite{Alloul:2013bka}.  
Here, and in all the following numerical calculations, 
we set $m_h=125$~GeV, $m_t=173$~GeV, $m_W=80.4$~GeV, 
$v=246$~GeV, $4\pi/e^2=128$ and $\sin^2\theta_W=0.233$ for 
the electroweak couplings.
Factorization scale is set at $\mu = (m_t + m_h)/4$ 
for the $ht$ and $h\bar{t}$ production via $t$-channel 
$W$ exchange processes in 5-flavor QCD, following 
Ref.\,\cite{Demartin:2015uha}.  
As for the QCD production of $ht\bar{t}$ processes, 
we set the factorization and renormalization scales 
both at $\mu=(2m_t+m_h)/2$, following 
Ref.\,\cite{Beenakker:2002nc}.
The QCD coupling at $\mu=m_Z$ is set at 
$\alpha_s(m_Z) = 0.118$\,\cite{Tanabashi:2018oca}. 

As is well known, the cross sections for the Higgs production 
with single $t$ or $\bar{t}$ are sensitive to the relative 
sign of the $htt$ and the $hWW$ couplings, which becomes 
13 times larger than the SM value at $|\xi_{htt}|=\pi$ 
where the sign of the $htt$ coupling is 
reversed\,\cite{Stirling:1992fx}.  
Because of this huge enhancement factor, 
LHC experiments\,\cite{Khachatryan:2015ota,CMS PAS HIG-17-005,CMS PAS HIG-17-009,CMS PAS HIG-17-016,Sirunyan:2018lzm} have ruled out 
the region around $|\xi_{htt}| \sim \pi$ for 
$\kappa_{htt}=1$. 
It is worth noting, however, that we focus our attention 
in this paper on a relatively small magnitude of the CP phase 
$|\xi_{htt}|\lesssim 0.1\pi$, where the total cross sections 
do not deviate much from the SM values, $\sigma(th+\bar{t}h)=60.85$\,fb and $\sigma(t\bar{t}h)=$ 406.26\,fb in the LO,  as shown in Fig.\,\ref{fig:cs}. 

Past studies of the $h$ and single $t$ or $\bar{t}$ 
production signal and backgrounds at hadron colliders 
include NLO corrections with the matching between the 4- and 5-flavor QCD predictions\,\cite{Demartin:2015uha, Maltoni:2001hu}, 
with Higgs decay channels $h\to WW/ZZ$\,\cite{Chang:2014rfa,Barger:2009ky}, 
$\gamma\gamma$\,\cite{Chang:2014rfa,Biswas:2012bd,Yue:2014tya,
Gritsan:2016hjl}, 
$b\bar{b}$\,\cite{Chang:2014rfa,Farina:2012xp,Agrawal:2012ga,Kobakhidze:2014gqa} and $\tau^+\tau^-$\,\cite{Chang:2014rfa}.  
CP phases of the top Yukawa couplings\,\cite{Atwood:2000tu} are studied in 
$t+h$ production~\cite{Biswas:2012bd,Yue:2014tya,Chang:2014rfa,Gritsan:2016hjl,Rindani:2016scj,Kraus:2019myc}, 
$ht\bar{t}$ production~\cite{Gritsan:2016hjl}, 
and in the loop induced vertices 
$hgg$ or $h\gamma\gamma$~\cite{Djouadi:2005gj}.
The first result of our study has been reported in\,\cite{Barger:2018tqn}, and related studies are found in\,\cite{Faroughy:2019ird}.

The paper is organized as follows.  
In section\,\ref{sec:HelAmp}, we give helicity amplitudes for all the 
four LO subprocesses analytically.  
In section\,\ref{sec:distributions}, we study event distributions of $ht$ 
and $h\bar{t}$ production with a tagged 
forward jet, and show the exchanged $W$ helicity 
decomposition in ${\tt Q}$ (the virtual $W$ mass)
and ${\tt W}$ (the invariant mass of the $th$ or $\bar{t}h$ 
system) distributions. 
In section\,\ref{sec:A_azimuthal}, we study the azimuthal angle asymmetry 
between the $W$ emission plane and the $W^+b \to th$ 
or $W^-\bar{b}\to \bar{t}h$ production plane about 
the $W$ momentum direction.  
In section\,\ref{sec:pol}, we study $t$ and $\bar{t}$ polarizations 
in the $t$ ($\bar{t}$) rest frames, as a function of ${\tt Q}$, ${\tt W}$ and 
the $W^+b\to th$ ($W^-\bar{b}\to \bar{t}h$) 
scattering angle $\theta^*$ in the $th$ ($\bar{t}h$) 
rest frame.  
In section\,\ref{section:Todd}, we study consequences of T and CP 
transformations, and show the 
possibility that CP violation signal can be distinguished 
from T-odd asymmetry arising from the final state 
scattering phase in $pp$ collisions, by measuring 
the $t$ and $\bar{t}$ polarizations perpendicular to 
the scattering plane. 
The last section\,\ref{sec:summary} gives a summary of our 
findings and remarks on possible measurements 
at HL-LHC.  
Appendix\,\ref{sec:AppA} gives the relation between the helicity amplitudes and $t$ and $\bar{t}$ spin polarizations, and Appendix\,\ref{sec:AppB} gives polarized $t$ and $\bar{t}$ decay distributions.

\section{Helicity Amplitudes} 
\label{sec:HelAmp}

\begin{figure}[h]
\begin{centering}
\begin{tabular}{c}
\includegraphics[width=0.35\textwidth]{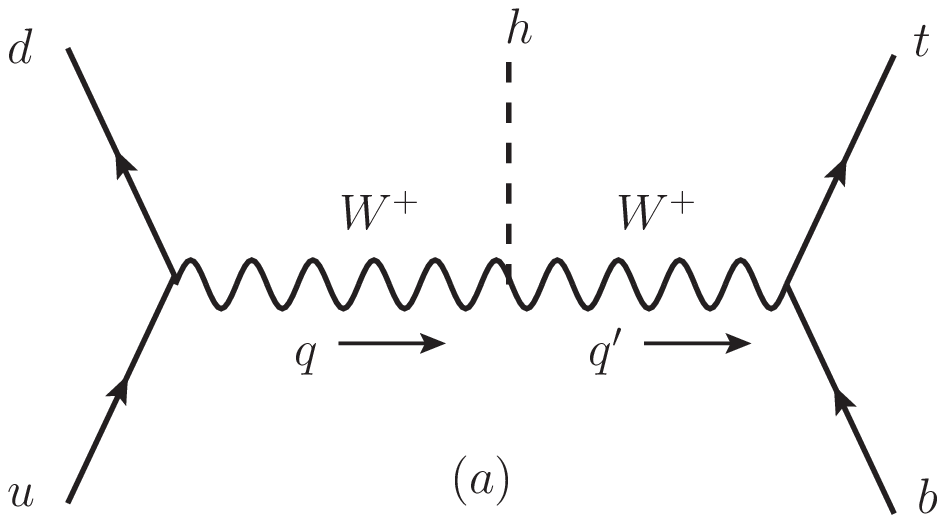}\label{fig:2a}
\includegraphics[width=0.35\textwidth]{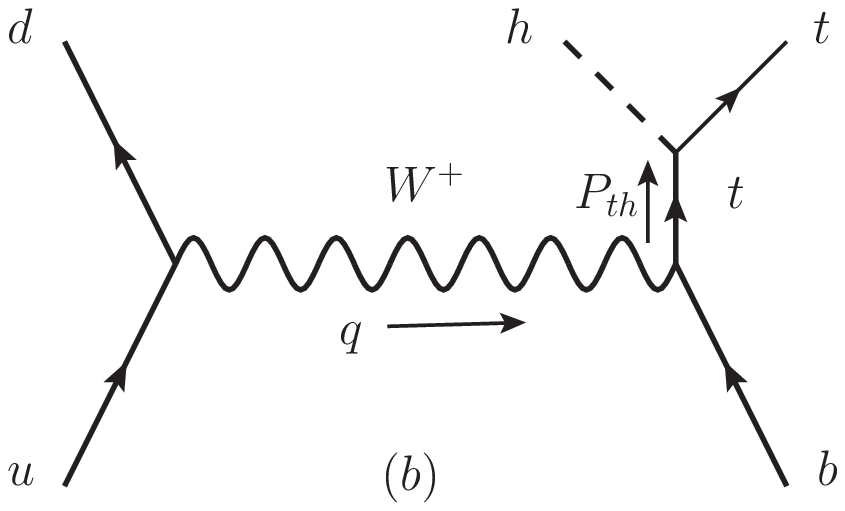}\label{fig:2b}
\end{tabular}
\caption {
Feynman diagrams of $ub\to dth$ subprocess. The four momenta $q^\mu$ and $q^{\prime \mu}$ along the $W^+$ and $P_{th}^\mu$ along the top propagators are shown with arrows.}
\label{fig:feyn}
\end{centering}
\end{figure}

In the SM, four subprocess contribute to single top plus 
Higgs production in the leading order
\begin{subequations}
\label{eq:sub}
\begin{align}
&ub\to dth\quad (cb\to sth) 
\label{eq:subub}\\ 
&\bar{d}b\to\bar{u}th \quad (\bar{s}b\to\bar{c}th)
\label{eq:subdxb}  
\end{align}
\label{eq:subtop}
\end{subequations}
and also to single anti-top plus Higgs production;
\begin{subequations}
\begin{align}
&d\bar{b}\to u\bar{t}h \quad (s\bar{b}\to c\bar{t}h)
\label{eq:subdbx}\\  
&\bar{u}\bar{b}\to \bar{d}\bar{t}h\quad (\bar{c}\bar{b}
\to \bar{s}\bar{t}h)
\label{eq:subuxbx} 
\end{align}
\label{eq:subtbar}
\end{subequations}

We work in 5-flavor QCD with massless $b$-quark 
distribution in the proton, where the matching with 
the 4-flavor QCD with massive $b$-quark has been 
shown for the single $t$ plus $h$ processes in the 
NLO level\,\cite{Beenakker:2002nc,Barger:2009ky}.
The subprocesses in the parenthesis with second generation 
quarks have exactly the same matrix elements when we ignore 
quark mass and CKM mixing effects. 

The Feynman diagrams of the subprocess 
$ub\to dth$ in Eq.(\ref{eq:subub}) are shown in 
Fig.~\ref{fig:feyn}. 
The left diagram~($a$) has the $hWW$ coupling, while 
the right diagram ($b$) has the $htt$ coupling. 
The $u\to dW^+$ emission part is common to both diagrams. 
The amplitudes for all the other subprocesses in 
Eq.\,(\ref{eq:sub}) are obtained by replacing the 
$u\to dW^+$ emission current by $c\to sW^+$, 
$\bar{d}\to\bar{u}W^+$ and $\bar{s}\to\bar{c}W^+$ current, 
respectively. 
The Feynman diagrams for anti-top plus Higgs production in 
Eq.\,(\ref{eq:subtbar}) are obtained simply by replacing 
the $W^+$ emission currents by the $W^-$ emission currents, 
and by reversing the fermion-number flow along 
the $b$ to $t$ transitions to make them $\bar{b}$ 
to $\bar{t}$ transitions.  

In $pp$ collisions, valence quark initiated subprocesses 
$ub\to dth$ (3a) and $d\bar{b}\to u\bar{t}h$ (4a) dominate 
the single top and anti-top production cross sections, 
respectively.
The amplitudes for the subprocess $ub\to dth$ in 
Fig.\,\ref{fig:feyn} are simply 
\begin{eqnarray}
{\cal M}_\sigma = -\frac{g^2}{2}
\bar{u}(p_d)\gamma_\mu\frac{1-\gamma_5}{2}
u(p_u)D_W^{\mu\nu}(q)\bar{u}(p_t,\sigma)
T_\nu \frac{1-\gamma_5}{2}u(p_b)
\label{eq:Msig}
\end{eqnarray}
with
\begin{eqnarray}
\label{eq:Tnu}
T^\nu = 
g_{htt}(\cos\xi+i\sin\xi\gamma_5)({\slashed P}_{th}+m_t) 
D_t(P_{th})\gamma^\nu
-g_{hWW}D_W^{\nu\rho}(q^\prime)\gamma_\rho
\end{eqnarray}
for the effective top Yukawa coupling of Eq.(\ref{eq:coupling}) 
and the SM $hWW$ coupling of Eq.\,(\ref{eq:ghww}). 
The propagator factors 
\begin{eqnarray}
\label{eq:DW(q)}
D_W^{\mu\nu}(q)=
\left(-g^{\mu\nu}+\frac{q^\mu q^\nu}{m_W^2}\right)D_W(q)
\end{eqnarray}
and $D_W^{\nu\rho}(q^\prime)$ are the $W$-propagators, 
with $D_W(q)=(q^2-m_W^2)^{-1}$, and 
$D_t(P_{th})=(P_{th}^2-m_t^2)^{-1}$ is for the top quark. 
The four momenta are depicted in Fig.\,\ref{fig:feyn} as
\begin{eqnarray}
q=p_u-p_d,\quad 
q^\prime = q - p_h = p_t-p_b,\quad 
P_{th}=q+p_b=p_t+p_h.
\label{eq:qqPth}
\end{eqnarray}
In the limit of neglecting all the quark masses except 
the top quark mass, $m_t$, the amplitudes depend only 
on the top quark helicity $\sigma/2$ for $\sigma=\pm1$, 
since only the left-handed quarks and right-handed 
anti-quarks contribute to the SM charged currents  
in the massless limit. 

Because the $W^+$ emission current of massless $u$ and $d$ 
quarks is conserved, only the spin 1 components of off-shell 
$W^+$ propagates in the common $D_W^{\mu\nu}(q)$ term 
in Eq.\,(\ref{eq:DW(q)}):  
\begin{eqnarray}
-g^{\mu\nu}+\frac{q^\mu q^\nu}{m_W^2}
\to 
-g^{\mu\nu}+\frac{q^\mu q^\nu}{q^2}
=
\sum_{\lambda=\pm1,0} (-1)^{\lambda+1}
\epsilon^\mu(q,\lambda)^\ast\epsilon^\nu(q,\lambda),
\label{eq:v-prop}
\end{eqnarray}
where $\lambda$ denotes the helicity of virtual $W^+$, 
and the $(-1)^{\lambda+1}$ factor appears for $q^2<0$. 
By replacing the covariant propagation factor in the 
common $W^+$ propagator with Eq.\,(\ref{eq:v-prop}), 
we can express the amplitudes Eq.\,(\ref{eq:Msig}) as 
a sum over the contributions of the three $W^+$ helicity 
states:
\begin{eqnarray}
{\cal M}_\sigma=\sum_{\lambda=\pm1,0}
{\cal J}_\lambda(u\to dW_\lambda^+)
{\cal T}_{\lambda\sigma}(W_\lambda^+b\to t_\sigma h)
\end{eqnarray}
with
\begin{eqnarray}
{\cal J}_\lambda=
\frac{g}{\sqrt{2}} D_W^{}(q)
\bar{u}(p_d) \gamma^\mu\frac{1-\gamma_5}{2}u(p_u)
\epsilon_\mu^\ast(q,\lambda)
(-1)^{\lambda}
\label{eq:Jlambda}
\end{eqnarray}
and
\begin{eqnarray}
\label{eq:Tlamsig}
{\cal T}_{\lambda\sigma}=
\frac{g}{\sqrt{2}} 
\bar{u}(p_t,\sigma)\left\{
g_{htt}(\cos\xi+i\sin\xi\gamma_t)D_t(P_{th})
(\slashed P_{th}+m_t)\gamma^\nu
-
g_{hWW}D_W^{\nu\rho}(q^\prime)\gamma_\rho\right\}
\frac{1-\gamma_5}{2}u(p_b)
\epsilon_\nu(q,\lambda).
\end{eqnarray}

\begin{figure}[t]
\begin{centering}
\begin{tabular}{c}
\includegraphics[width=0.55\textwidth]{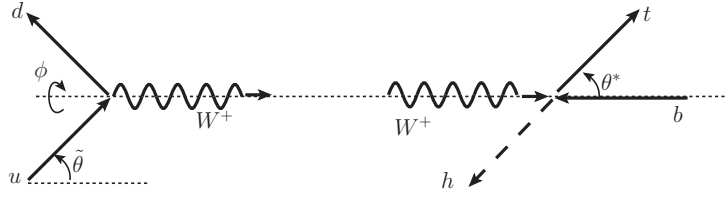}
\end{tabular}
\caption {Scattering angles $\tilde{\theta}$, $\phi$ and $\theta^\ast$. The polar angle $\tilde{\theta}$ is defined in the Breit frame, whereas $\theta^\ast$ is defined in the $W^+b$ rest frame, for the common polar axis along the $W$ momentum direction. The azimuthal angle $\phi$ is the angle between the emission plane and the scattering plane.}
\label{fig:frame}
\end{centering}
\end{figure}

We calculate the helicity amplitudes ${\cal T}_{\lambda\sigma}$ 
for $W^+b\to th$ process in the $th$ or $W^+b$ rest frame. 
Therefore all the polarization asymmetries presented below 
refer to the top quark helicity in the $th$ rest frame, 
see Fig.\,\ref{fig:frame}. 
On the other hand, because massless quark helicities are 
Lorentz invariant, and the $W^+$ helicity is boost 
invariant along the $W^+$ momentum direction, which we 
take as the polar axis in Fig.\,\ref{fig:frame}, 
we can evaluate the $u\to dW^+$ emission amplitudes 
in the Breit frame~\cite{Hagiwara:2009wt}, where the 
$W^+$ four momentum has only the helicity axis component
\begin{eqnarray}
q^\mu=(0,0,0,{\tt Q})
 \end{eqnarray} 
with ${\tt Q} >0$ and ${\tt Q}^2=-q^2$. 
The $u$ and $d$ quark four momenta are 
 \begin{subequations}
 \label{eq:pud}
\begin{align}
p_u^\mu&=\tilde{\omega}(1,\sin\tilde{\theta}\cos\phi,-\sin\tilde{\theta}\sin\phi,\cos\tilde{\theta}),\\
p_d^\mu&=\tilde{\omega}(1,\sin\tilde{\theta}\cos\phi,-\sin\tilde{\theta}\sin\phi,-\cos\tilde{\theta}),
\end{align}
 \end{subequations}
where their common energy $\tilde{\omega}$ and the reflecting 
momentum along the polar axis are, respectively,
 \begin{subequations}
 \begin{align}
 \tilde{\omega}&=({\tt Q}/2)\left[2\hat{s}/({\tt W}^2+{\tt Q}^2)-1\right],\\
 \tilde{\omega}&\cos\tilde{\theta}={\tt Q}/2,
 \end{align}
 \end{subequations}
with $\hat{s}=(p_u+p_b)^2$ and 
${\tt W}=\sqrt{P_{th}^2}=\sqrt{(p_t+p_h)^2}$. 
In Eq.\,(\ref{eq:pud}) and in Fig.\,\ref{fig:frame}, the $u\to dW^+$ emission plane is rotated by 
$-\phi$ about the $z$-axis, so that the top quark azimuthal 
angle measured from the $u\to dW^+$ emission plane is $\phi$.
The $u\to dW^+$ emission amplitudes have very compact 
and intuitive expressions in the Breit frame: 
\begin{subequations}
\begin{align}
{\cal J}_\pm&=\frac{g}{\sqrt{2}}D_W(q)(2\tilde{\omega})
\left(e^{\pm i\phi}
\frac{1\mp\cos\tilde{\theta}}{\sqrt{2}}\right)
\\
{\cal J}_0&=\frac{g}{\sqrt{2}}D_W(q)
(2\tilde{\omega})\sin\tilde{\theta}.  
\end{align}
\label{eq:J}
\end{subequations}
Here we adopt 
\begin{eqnarray}
\epsilon^\mu(q,\lambda=\pm1)=\frac{1}{\sqrt{2}}(0,1,\pm i,0),
\quad\quad 
\epsilon^\mu(q,\lambda=0)=(1,0,0,0),
\label{eq:epsilon}
\end{eqnarray}
for the three polarization vectors, which differs by the sign of 
the $\lambda=+1$ vector from the standard Jacob-Wick 
convention. 
The convention dependence cancels in the product, and our 
choice makes CP transformation properties of the 
sub-amplitudes, ${\cal J}_\lambda$ and ${\cal T}_{\lambda\sigma}$, 
simple because 
\begin{eqnarray}
\epsilon^\mu(q,\lambda)^\ast=\epsilon^\mu(q,-\lambda).
\end{eqnarray}
It is interesting to note~\cite{Hagiwara:2009wt} that 
the $u\to dW^+$ emission amplitudes can be expressed 
in terms of Wigner's $d$-functions.
In terms of the invariants, they are expressed as 
\begin{subequations}
\begin{align}
(2\tilde{\omega})\frac{1+\cos\tilde\theta}{2}
&=\frac{{\tt Q}}{1-x+{\tt Q}^2/\hat{s}}
\\
(2\tilde{\omega}) \frac{\sin\tilde{\theta}}{\sqrt{2}}
&=\frac{{\tt Q}}{1-x+{\tt Q}^2/\hat{s}}\sqrt{x-{\tt Q}^2/\hat{s}}
\\
(2\tilde{\omega})(1-\cos\tilde{\theta})
&=\frac{{\tt Q}}{1-x+{\tt Q}^2/\hat{s}}\left(x-{\tt Q}^2/\hat{s}\right)
\end{align}
\end{subequations}
where 
\begin{eqnarray}
x=1-{\tt W}^2/\hat{s}
\end{eqnarray}
is the energy fraction of the $d$-quark 
in the $ub$ collision rest frame. 
It should be noted that for typical events with 
$x\lesssim 0.1$, the ordering
\begin{eqnarray}
\frac{1+\cos\tilde{\theta}}{2}
\gg \frac{\sin\tilde{\theta}}{\sqrt{2}}
\gg \frac{1-\cos\tilde{\theta}}{2}
\end{eqnarray}
holds among the magnitudes of the $d$-functions. 
In particular, $J_-$ for the helicity $\lambda=-1$ 
$W^+$ dominates over $J_+$, because left-handed 
$u$-quark tends to emit a left-handed $W$-boson 
in the forward direction. 

The helicity amplitudes ${\cal T}_{\lambda\sigma}$ for 
$W^+_{\lambda}b\to t_\sigma h$ process are calculated 
in the $th$ rest frame. 
We first express ${\cal T}_{\lambda\sigma}$ (\ref{eq:Tlamsig}) in terms of chiral
two-spinors~\cite{Hagiwara:1985yu}
\begin{eqnarray}
{\cal T}_{\lambda\sigma}
&=&
\frac{g}{\sqrt{2}} g_{htt}^{}D_t(P_{th})
\left[%
e^{-i\xi}u_R^\dagger(p_t,\sigma)~P\cdot\sigma_+%
+
e^{i\xi}m_tu_L^\dagger(p_t,\sigma)
\right]%
~\epsilon(q,\lambda)\cdot\sigma_- ~ u_L(p_b)
\nonumber\\
&+& \frac{g}{\sqrt{2}} g_{hWW}^{}D_W(q^\prime)
\left[%
u_L^\dagger(p_t,\sigma)
~\epsilon(q,\lambda)\cdot\sigma_- 
+
\frac{m_t}{m_W^2}~p_b\cdot\epsilon(q,\lambda)%
~u_R^\dagger(p_t,\sigma)
 \right]
u_L(p_b)
\end{eqnarray}
where we denote $P=P_{th},~ \xi=\xi_{htt},$ and 
$\sigma_{\pm}^\mu=(1,\pm\vec{\sigma})$ are the chiral 
four-vectors of $\sigma$ matrices. 
We note that the chirality flip term for the right-handed 
top with $e^{-i\xi}$ phase factor grows with $P$, 
while the chirality non-flip term for the left-handed 
top with $e^{i\xi}$ is proportional to $m_t$, 
because of the chirality flip by the Yukawa interactions. 
As for the $W$-exchange amplitudes, the chirality flip 
right-handed top proportional to $m_t$ is non-negligible 
because of the scalar component of the exchanged $W$ boson, 
which has the $1/m_W^2$ factor.
In the $th$ rest frame, where the $W^+$ momentum is along 
the positive $z$-axis, the four momenta are given by 
\begin{subequations}
\begin{align}
q&=\frac{\mathtt W}{2}
\left(
1-\frac{{\tt Q}^2}{ {\tt W}^2},~0,~0,~1+\frac{{\tt Q}^2}{\tt W^2}
\right)
=
(q^{0\ast},~0,~0,~q^\ast)
\label{eq:qmomentum}
\\
p_b&=\frac{\mathtt W}{2}
\left(
1+\frac{ {\tt Q}^2}{{\tt W}^2},~0,~0,
~-\left(1+\frac{{\tt Q}^2}{{\tt W}^2}\right)
\right)
=
(q^{\ast},~0,~0,~-q^\ast)
\\
p_t&=\frac{\mathtt W}{2}
\left(1+\frac{m_t^2-m_h^2}{{\tt W}^2},
~\bar{\beta}\sin\theta^\ast,~0,
~\bar{\beta}\cos\theta^\ast\right)
=
(E_t^\ast,~p^\ast\sin\theta^\ast,~0,~p^\ast\cos\theta^\ast)
\\
p_h&=\frac{\mathtt W}{2}
\left(1+\frac{m_h^2-m_t^2}{{\tt W}^2},
~-\bar{\beta}\sin\theta^\ast,~0,
~-\bar{\beta}\cos\theta^\ast\right)
=
(E_h^\ast,~-p^\ast\sin\theta^\ast,~0,~-p^\ast\cos\theta^\ast)
\end{align}\label{eq:qpbptph1}
\end{subequations}
\hspace{-0.2cm}
where 
$\bar{\beta} = 2p^\ast/{\tt W} = 
\left[({\tt W}+m_t+m_h)({\tt W}+m_t-m_h)({\tt W}-m_t+m_h)
({\tt W}-m_t-m_h)\right]^{1/2}/{\tt W}^2$ 
is the c.m. momentum $p^\ast$ of $t$ and $h$ in the $th$ 
rest frame in units of ${\tt W}/2$. 
The amplitudes ${\cal T}_{\lambda\sigma}$ can be calculated 
straightforwardly, giving 
%
%
\begin{subequations}
\begin{align}
{\cal T}_{+\pm}&
=\pm \frac{g}{\sqrt{2}} g_{hWW}^{}D_W(q^\prime)
\frac{mp^\ast}{m_W^2}
\sqrt{q^\ast(E^\ast\pm p^\ast)(1\pm\cos\theta^\ast)}
\frac{\sin\theta^\ast}{\sqrt{2}}
\\
{\cal T}_{-\pm}&
= \frac{g}{\sqrt{2}} g_{hWW}^{}D_W(q^\prime)
\left[\pm\frac{mp^\ast}{m_W^2}
\sqrt{q^\ast(E^\ast\pm p^\ast)(1\pm\cos\theta^\ast)}
\frac{\sin\theta^\ast}{\sqrt{2}}
+\sqrt{2q^\ast(E^\ast\mp p^\ast)(1\mp\cos\theta^\ast)}\right]
\nonumber\\
&~ ~+\frac{g}{\sqrt{2}} g_{htt}^{}D_t(P)
\left[\left(e^{-i\xi}{\tt W}\sqrt{E^\ast\pm p^\ast}
+e^{i\xi}m\sqrt{E^\ast\mp p^\ast}\right)
\sqrt{2q^\ast(1\mp\cos\theta^\ast)}
\right]
\\
{\cal T}_{0\pm}&
= \pm \frac{g}{\sqrt{2}} g_{hWW}^{}D_W(q^\prime)
\left[
\left(
\frac{m(q^\ast E_h^\ast+q^{0\ast}p^\ast\cos\theta^\ast)}
{m_W^2{\tt Q}}\sqrt{E^\ast\pm p^\ast}
+\frac{\tt W}{{\tt Q}}\sqrt{E^\ast\mp p^\ast}
\right)
\sqrt{q^\ast(1\pm\cos\theta^\ast)}
\right]
\nonumber\\
&~~ \pm\frac{g}{\sqrt{2}} g_{htt}D_t(P)
\frac{{\tt W}}{{\tt Q}}
\left[\left(e^{-i\xi}{\tt W}\sqrt{E^\ast\pm p^\ast}
+e^{i\xi}m\sqrt{E^\ast\mp p^\ast}\right)
\sqrt{q^\ast(1\mp\cos\theta^\ast)}
\right]\,\noindent
\label{eq:T0}
\end{align}
\label{eq:T}
\end{subequations}
\hspace{-0.15cm}where we denote $m=m_t$ and $E^\ast=E^\ast_t$. 
Note that the term $\sqrt{E^\ast+p^\ast}$ appears when 
the top helicity matches its chirality, while 
$\sqrt{E^\ast-p^\ast}$ when they mismatch. 
The amplitude for $\lambda=+1$ does not have the top Yukawa 
coupling contribution because the angular momentum along 
the $z$-axis is $J_z=+\frac{3}{2}$ for the left-handed 
$b$-quark, which cannot couple to $J=1/2$ top quark. 
For $\lambda=-1$ and $\lambda=0$ $W^+$, both $W$ and $t$ 
exchange amplitudes contribute. 
Most importantly, the $\lambda=0$ amplitudes are enhanced 
by the factor of ${\tt W/Q}$, which is a consequence of 
the boost factor of the longitudinally polarized 
$\lambda=0$ $W^+$ wave function. 
The polarization vectors in Eq.\,(\ref{eq:epsilon}) in 
the Breit frame are invariant for $\lambda=\pm1$, but the longitudinal vector becomes
\begin{eqnarray}
\epsilon^\mu(q,\lambda=0)=\frac{1}{\tt Q}(q^\ast,~0,~0,~q^{0\ast})
\end{eqnarray}
in the $th$ rest frame, where both $q^\ast$ and $q^{0\ast}$ 
are the order of ${\tt W}/2$ as in Eq.\,(\ref{eq:qmomentum}).

Summing over the three $W$ polarization contributions, 
we find the amplitudes\,\cite{Barger:2018tqn} 
\begin{subequations}
\label{eq:amp}
{\allowdisplaybreaks
\begin{align}
{\cal M}_+^{}&
=\frac{1-\tilde{c}}{2}e^{i\phi}\sin\frac{\theta^\ast}{2}
\left[\frac{1+\cos\theta^\ast}{4}\bar{\beta}A\right]
\nonumber\\
&+\frac{1+\tilde{c}}{2}e^{-i\phi}\sin\frac{\theta^\ast}{2}
\Bigg[\left(\frac{1+\cos\theta^\ast}{4}\bar{\beta}
+\epsilon\delta\delta^\prime\right)A
-\left(e^{-i\xi}+\delta\delta^\prime e^{i\xi}\right)B\Bigg]
\nonumber\\
&+\frac{\tilde{s}}{2}\frac{\tt W}{{\tt Q}}\cos\frac{\theta^\ast}{2}
\Bigg[\left(\frac{q^\ast E_h^\ast
+q^{0\ast}p^\ast\cos\theta^\ast}{{\tt W}^2}
+\epsilon\delta\delta^\prime\right)A
-\left(e^{-i\xi}+\delta\delta^\prime e^{i\xi}\right)B\Bigg],
\label{eq:M+}\\
{\cal M}_-^{}&
=-\frac{1-\tilde{c}}{2}e^{i\phi}\cos\frac{\theta^\ast}{2}
\left[\frac{1-\cos\theta^\ast}{4}\bar{\beta} A\right]\delta
\nonumber\\
&-\frac{1+\tilde{c}}{2}e^{-i\phi}\cos\frac{\theta^\ast}{2}
\Bigg[\left(\frac{1-\cos\theta^\ast}{4}\bar{\beta}
-\epsilon\frac{\delta^\prime}{\delta}\right)A
+\left( e^{-i\xi}+\frac{\delta^\prime}{\delta}e^{i\xi}
\right )B\Bigg]\delta
\nonumber\\
&-\frac{\tilde{s}}{2}\frac{\tt W}{{\tt Q}}\sin\frac{\theta^\ast}{2}
\Bigg[\left(
\frac{q^\ast E_h^\ast+q^{0\ast}p^\ast\cos\theta^\ast}
{{\tt W}^2}+\epsilon\frac{\delta^\prime}{\delta}\right)A
-\left(e^{-i\xi}+\frac{\delta^\prime}{\delta} e^{i\xi}
\right)B\Bigg]{\delta}.
\label{eq:M-}
\end{align}}
\end{subequations}
\hspace{-0.15cm}where the factors 
\begin{subequations}\label{eq:AB}
\begin{align}
A&=2g^2g_{hWW}^{}\frac{m{\tt W}}{m_W^2}D_W^{}(q)D_W(q^\prime)
\tilde{\omega}\sqrt{2q^\ast(E^\ast+p^\ast)},
\label{eq:A}\\
B&=-2g^2g_{htt}^{}{\tt W}D_W^{}(q)D_t^{}(P)\tilde{\omega}
\sqrt{2q^\ast (E^\ast+p^\ast)},
\label{eq:B}
\end{align}
\end{subequations}
are chosen such that they are positive definite. 
The $\epsilon$, $\delta$, and $\delta^\prime$ factors are
\begin{eqnarray}
\label{eq:deltadeltaprime}
\epsilon=\frac{m_W^2}{m^2},\quad 
\delta=\frac{m}{E^\ast+p^\ast},\quad 
\delta^\prime=\frac{m}{\tt W},
\end{eqnarray}
where $\epsilon \simeq 0.21$, $\delta$ and $\delta^\prime$ 
are all small at large ${\tt W}$, and in particular, 
$\delta\simeq \delta^\prime$ holds rather accurately  
except in the vicinity of $th$ production threshold, 
${\tt W}\simeq m_t+m_h$. 
At ${\tt W}\gtrsim 400$ GeV, the amplitudes are well 
approximated as 
\begin{subequations}\label{eq:simMpm}
\begin{align}
{\cal M}_+&\sim
\left[\frac{1+\tilde{c}}{2}e^{-i\phi}\sin\frac{\theta^\ast}{2}
+
\frac{\tt W}{{\tt Q}}\frac{\tilde{s}}{2}\cos\frac{\theta^\ast}{2}
\right]
\left[\frac{1+\cos\theta^\ast}{4}A-e^{-i\xi}~B\right]
,\\
{\cal M}_-&\sim
-\frac{1+\tilde{c}}{2}\cos\frac{\theta^\ast}{2}e^{-i\phi}
\left[
\left(\frac{1-\cos\theta^\ast}{4}-\epsilon\right)A+2\cos\xi B
\right]\delta
-\frac{\tt W}{{\tt Q}}\frac{\tilde{s}}{2}
\sin\frac{\theta^\ast}{2}\left[\left(
\frac{1+\cos\theta^\ast}{4}+\epsilon\right)A-2\cos\xi B
\right]\delta,
\label{eq:approM-}
\end{align}
\end{subequations}
where we have dropped $\lambda=+1$ contributions which are 
suppressed at high ${\tt W}/{\tt Q}$. 
The above approximations show that the leading 
$\lambda=0$ contributions with the ${\tt W}/{\tt Q}$ enhancement factor 
are proportional
\begin{subequations}
\begin{align}
&\frac{1+\cos\theta^\ast}{4}A-e^{-i\xi}~B 
\quad {\rm for} ~{\cal M}_+,
\\
&\left(\frac{1+\cos\theta^\ast}{4}+\epsilon\right)A
-2\cos\xi ~B \quad {\rm for} ~{\cal M}_-.
\end{align}
\end{subequations}
Because both $A$ and $B$ terms are positive definite, their
 magnitudes are smallest at $\xi=0$ (SM), 
where the $W$ exchange term $A$ and the $t$-exchange term 
$B$ interfere destructively, whereas they become largest 
at $|\xi|=\pi$ where the two terms interfere constructively, 
explaining the order of magnitude difference in the total 
cross section between $\xi=0$ and $|\xi|=\pi$ shown in 
Fig.\,\ref{fig:cs}. 
This strong interference between the two amplitudes gives 
the opportunity to accurately measure the $htt$ Yukawa 
coupling with respect to the $hWW$ coupling. 

Another important observation from the above approximation 
is that the CP-violation (CPV) effects proportional to 
$\sin\xi$ are significant only in the amplitude of the 
right-handed helicity top quark, ${\cal M}_+$, because ${\cal M}_-$ is proportional to $e^{-i\xi}+e^{+i\xi}=2\cos\xi$ at large ${\tt W}$ where $\delta=\delta^\prime$.
We note here that ${\cal M}_+$ is the leading helicity 
amplitude at large ${\tt W}$, where the chirality flip 
Yukawa interactions give right-handed top quark from the 
left-handed $b$-quark. 
The negative helicity amplitudes ${\cal M}_-$ is suppressed 
by an additional chirality flip of the top quark, 
indicated by the factor $\delta=m/{\tt W}$ in 
Eq.\,(\ref{eq:approM-}).

Before starting discussions about signals at the LHC, let us 
complete all the helicity amplitudes of the contributing 
subprocess for both $th$ and $\bar{t}h$ productions. 
First, the amplitudes for the subprocesses $cb\to sth$ are 
the same as those for $ub\to dth$ in our approximation of 
neglecting quark masses and CKM mixing: 
\begin{eqnarray}
\label{eq:Mub=cb}
{\cal M}_\sigma(ub\to dth)
={\cal M}_\sigma (cb\to sth) 
=\sum_\lambda {\cal J}_\lambda {\cal T}_{\lambda\sigma}
\end{eqnarray}
as summarized in Eqs.\,(\ref{eq:amp}). 
There are two additional contributions to $th$ production
from the anti-quark distributions of proton
\begin{eqnarray}\label{eq:Mdbarb=sbarb}
{\cal M}_\sigma (\bar{d}b\to\bar{u}th) 
={\cal M}_\sigma(\bar{s}b\to\bar{c}th)
=\sum_\lambda\overline{{\cal J}}_\lambda{\cal T}_{\lambda\sigma}
\end{eqnarray}
where the $\bar{d}\to\bar{u}W^+$ emission amplitudes are 
\begin{eqnarray}
\overline{{\cal J}}_\lambda=\frac{g}{\sqrt{2}}
\bar{v}(p_{\bar{d}}^{})\gamma^\mu\frac{1-\gamma_5}{2}
v(p_{\bar{u}}^{})\epsilon_\mu^\ast(q,\lambda)(-1)^\lambda
\end{eqnarray}
In the Breit frame, they are expressed as
\begin{subequations}
\begin{align}
\overline{\cal J}_{\pm}&
=\frac{g}{\sqrt{2}}D_W(q)(2\tilde{\omega})
\left(e^{\pm i\phi}\frac{1\pm\cos\tilde{\theta}}{\sqrt{2}}
\right),\\
\overline{\cal J}_0&
=\frac{g}{\sqrt{2}}D_W(q)\left(2\tilde{\omega}\right)
\sin\tilde{\theta}.
\end{align}
\label{eq:Jbar}
\end{subequations}
We note the relation 
\begin{eqnarray}\label{eq:JbarVSJbar*}
\overline{{\cal J}}_\lambda(\tilde{\theta},\phi)
=
{{\cal J}}_{-\lambda}(\tilde{\theta},-\phi)
=
{\cal J}_{-\lambda}^\ast(\tilde{\theta},\phi)
\end{eqnarray}
between ${{\cal J}}_\lambda$ and $\overline{{\cal J}}_\lambda$. 
The matrix elements for the $W^+$ emission from anti-quarks differ from those from quarks by simply replacing 
$1\pm\tilde{c}$ by $1\mp \tilde{c}$, thereby changing the 
preferred helicity of $W^+$ from $\lambda=-1$ 
(for $u\to dW^+$ and $c\to sW^+$) to 
$\lambda=+1$ (for $\bar{d}\to \bar{u}W^+$ and 
$\bar{s}\to \bar{c}W^+$). 
The $\lambda=0$ amplitude remains the same. 
Note that our special phase convention for the vector 
boson polarization vectors in Eq.\,(\ref{eq:epsilon}) allows the identities 
in Eq.\,(\ref{eq:JbarVSJbar*}) to hold without the $\lambda$-dependent 
sign factor, $(-1)^\lambda$, that appears in the standard Jacob-Wick convention.    

Now the $\bar{t}h$ production amplitudes are 
\begin{subequations}
\begin{align}
\overline{\cal M}_{\bar{\sigma}}(d\bar{b}\to u\bar{t}h)
&=
\overline{\cal M}_{\bar{\sigma}}(s\bar{b}\to c\bar{t}h)
=
\sum_{\lambda}
J_{\lambda} \overline{\cal T}_{\lambda\bar{\sigma}}
\label{eq:Mbar_dbbar}
\\
\overline{\cal M}_{\bar{\sigma}}
(\bar{u}\bar{b}\to \bar{d}\bar{t}h)
&=
\overline{\cal M}_{\bar{\sigma}}
(\bar{c}\bar{b}\to \bar{s}\bar{t}h)
=
\sum_{\lambda} 
\overline{{\cal J}}_{\lambda} \overline{\cal T}_{\lambda\bar{\sigma}}
\label{eq:Mbar_ubarbbar}
\end{align}
\end{subequations}
where ${\cal J}_\lambda$ and $\overline{{\cal J}}_\lambda$ are the same as in 
Eqs.(\ref{eq:J}) and (\ref{eq:Jbar}), respectively, and 
the $W^-\bar{b}\to\bar{t}h$ amplitudes 
$\overline{\cal T}_{\lambda\bar{\sigma}}$ are obtained 
from ${\cal T}_{\lambda\sigma}$ by CP transformation
\begin{eqnarray}
\overline{\cal T}_{\lambda\bar{\sigma}}(\theta^\ast,\xi)
=
-\bar{\sigma}{\cal T}_{-\lambda,-\bar{\sigma}}(\theta^\ast,-\xi)
=
-\bar{\sigma}{\cal T}_{-\lambda,-\bar{\sigma}}^\ast (\theta^\ast,\xi)
\label{eq:CPtransform}
\end{eqnarray}
Note that the first identity above tells the invariance 
of the amplitudes when all the initial and final states 
are CP transformed, 
along with the reversal of the sign of the CP phase. 
The latter equality is valid in our tree-level expressions 
Eqs.\,(\ref{eq:T}) where absorptive parts of the amplitudes 
(including the top quark width) are set to be zero.
 \footnote{
 Note that the sign factor, $-\bar{\sigma}$, in the identities\,(\ref{eq:CPtransform}) is a consequence of the phase convention of  Ref.\,\cite{Hagiwara:1985yu,Murayama:1992gi} where the $v$-spinors for anti-fermions are expressed as
 \begin{eqnarray}
&&v_L(p,\bar{\sigma})=i\sigma^2u_R(p,\bar{\sigma})^\ast=-\bar{\sigma}\sqrt{E+\bar{\sigma}p}~\chi_{-\bar{\sigma}}(\vec{p})
 \nonumber\\
 && v_R(p,\bar{\sigma})=-i\sigma^2u_L(p,\bar{\sigma})^\ast=\bar{\sigma}\sqrt{E-\bar{\sigma}p}~\chi_{-\bar{\sigma}}(\vec{p}).\,
 \nonumber
 \end{eqnarray}
 See also Appendix.\,B of Ref.\,\cite{Hagiwara:2017ban}.
 }

It is instructive to compare the amplitudes of the two 
subprocesses which are related by CP transformation, 
such as between the amplitudes\,(\ref{eq:amp}) or\,(\ref{eq:Mub=cb}) for 
$     u      b  \to d            t h$ 
and those of Eq.\,(\ref{eq:Mbar_ubarbbar}) for 
$\bar{u}\bar{b} \to \bar{d}\bar{t}h$,
\begin{subequations}
\label{eq:ub-vs-uxbx}
\begin{align}
{\cal M}_{\sigma}(ub\to dth;
\tilde{\theta},\phi,\theta^\ast;\xi)
&=
\sum_\lambda {\cal J}_\lambda(\tilde{\theta},\phi) 
{\cal T}_{\lambda \sigma}(\theta^\ast,\xi) ,
\label{eq:ub2dth}
\\
\overline{\cal M}_{\bar{\sigma}}
(\bar{u}\bar{b}\to\bar{d}\bar{t}h;
\tilde{\theta},\phi,\theta^\ast;\xi)
&=
\sum_\lambda \overline{{\cal J}}_{\lambda}(\tilde{\theta},\phi) 
\overline{\cal T}_{\lambda \bar{\sigma}}(\theta^\ast,\xi).
\label{eq:uxbx2dxtxh} 
\end{align}
\end{subequations}
By using the identities\,(\ref{eq:JbarVSJbar*}) and\,(\ref{eq:CPtransform}) among the 
sub-amplitudes, we find the relation
\begin{eqnarray}
\overline{\cal M}_{-\sigma}
(\bar{u}\bar{b}\to\bar{d}\bar{t}h;
\tilde{\theta},\phi,\theta^\ast;\xi)
=
\sigma{\cal M}_{\sigma}(ub\to dth;
\tilde{\theta},-\phi,\theta^\ast;-\xi),
\label{eq:cp-trans}
\end{eqnarray}
between $\overline{{\cal M}}_\sigma$ and ${\cal M}_\sigma$, when $\bar{\sigma}=-\sigma$.
It is worth noting that if we ignore the absorptive phase 
of the amplitudes, such as the top quark width in the 
propagator, the above identity gives 
\begin{eqnarray}
\overline{\cal M}_{-\sigma}
(\bar{u}\bar{b}\to\bar{d}\bar{t}h;
\tilde{\theta},\phi,\theta^\ast;\xi)
=
\sigma{\cal M}_{\sigma}^\ast
(ub\to dth;\tilde{\theta},\phi,\theta^\ast;\xi),  
\label{eq:cptn}
\end{eqnarray}
because both $\phi$ and $\xi$ appear in the amplitudes 
only as the phase factor, $e^{\pm i\phi}$ and $e^{\pm i\xi}$.  
This tells that all the distributions 
of the CP transformed processes are identical even in 
the presence of CP-violating phase, $\xi\neq 0$, if we 
ignore the absorptive amplitudes from the final state interactions.  
We will discuss the origin of this somewhat unexpected 
property among the amplitudes in section\,\ref{sec:pol}.  

In $pp$ collisions, the dominant subprocess for single 
production of Higgs and anti-top quark comes from the 
collision of valence down-quark scattering with $\bar{b}$ 
quark, whose amplitudes are given by Eq.\,(\ref{eq:Mbar_dbbar}).  
Since the properties of the $\bar{t}h$ production 
processes are governed by these amplitudes, we give 
their explicit form by using the same $A$ and $B$ factors 
of Eq.\,(\ref{eq:AB}): 
%
\begin{subequations}\label{eq:amptbar}
{\allowdisplaybreaks
\begin{align}
\overline{{\cal M}}_+^{}
&=\frac{1-\tilde{c}}{2}e^{i\phi}\cos\frac{\theta^\ast}{2}
\Bigg[\left(\frac{1-\cos\theta^\ast}{4}\bar{\beta}
-\epsilon\frac{\delta^\prime}{\delta}\right)A
+\left( e^{i\xi}+\frac{\delta^\prime}{\delta}e^{-i\xi}\right)
B\Bigg]\delta
\nonumber\\
&+\frac{1+\tilde{c}}{2}e^{-i\phi}\cos\frac{\theta^\ast}{2}
\frac{1-\cos\theta^\ast}{4}\bar{\beta} A\delta
\nonumber\\
&+\frac{\tilde{s}}{2}\frac{\tt W}{{\tt Q}}\sin\frac{\theta^\ast}{2}
\Bigg[\left(
\frac{q^\ast E_h^\ast+q^{0\ast}p^\ast\cos\theta^\ast}
{{\tt W}^2}+\epsilon\frac{\delta^\prime}{\delta}\right)A
-\left(e^{i\xi}+\frac{\delta^\prime}{\delta}e^{-i\xi}\right)
B\Bigg]{\delta}.
\label{eq:M+tbar}\\
\nonumber\\
\overline{{\cal M}}_-^{}
&=\frac{1-\tilde{c}}{2}e^{i\phi}\sin\frac{\theta^\ast}{2}
\Bigg[\left(\frac{1+\cos\theta^\ast}{4}\bar{\beta}
+\epsilon\delta\delta^\prime\right)A
-\left(e^{i\xi}+\delta\delta^\prime e^{-i\xi}\right)
B\Bigg]
\nonumber\\
&+\frac{1+\tilde{c}}{2}e^{-i\phi}\sin\frac{\theta^\ast}{2}
\frac{1+\cos\theta^\ast}{4}\bar{\beta}A
\nonumber\\
&+\frac{\tilde{s}}{2}\frac{\tt W}{{\tt Q}}\cos\frac{\theta^\ast}{2}
\Bigg[\left(
\frac{q^\ast E_h^\ast+q^{0\ast}p^\ast\cos\theta^\ast}
{{\tt W}^2}+\epsilon\delta\delta^\prime\right)A
-\left(e^{i\xi}+\delta\delta^\prime e^{-i\xi}\right)B\Bigg]\,
\label{eq:M-tbar}
\end{align}}
\end{subequations}
\hspace{-0.15cm}where the $d\to uW^-$ emission amplitudes $J_\lambda$ are 
the same as the $u\to dW^+$ emission amplitudes in the 
$ub\to dth$ subprocess amplitudes, Eq.\,(\ref{eq:amp}), 
while the $W^-\bar{b}\to\bar{t}h$ amplitudes 
$\overline{{\cal T}}_{\lambda\bar{\sigma}}$ are obtained from 
the $W^+b\to th$ amplitudes 
${\cal T}_{\lambda\sigma}$ by CP transformation in 
Eq.\,({\ref{eq:CPtransform}}). 
The chirality favored helicity of $\bar{t}$ from 
right-handed $\bar{b}$ is now $-1/2$, and the 
corresponding amplitude $\overline{{\cal M}}_-^{}$ 
has the leading $e^{i\xi}$ factor from the 
$t_L^\dagger t_R$ term in the Lagrangian Eq.\,(\ref{eq:coupling}), 
while the contribution of the 
$e^{-i\xi}t_R^\dagger t_L$ term is doubly chirality 
suppressed, by the $\delta\delta'$ factor. 
In the helicity $+1/2$ amplitude $\overline{{\cal M}}_+^{}$,  
single chirality flip (in addition to the flip due 
to the Yukawa interaction) is necessary, either in 
the spinor wave function (giving $\delta$), or in  
the top quark propagator (giving $\delta'$).  
Summing up, we find $\overline{{\cal M}}_-^{}$ to have 
significant imaginary part proportional to $\sin\xi$, 
whereas $\overline{{\cal M}}_+^{}$ is almost 
proportional to $\cos\xi$, which are opposite of 
what we find for the single $t$ and $h$ production 
amplitudes.  
%
%
\section{Differential Cross Sections}
\label{sec:distributions}
The differential cross section in $pp$ collisions from the subprocess $ub\to dth$ is given at leading order by
\begin{eqnarray}
\label{eq:xsdiff}
d\sigma=\int dx_1 \int dx_2 \left[D_u(x_1,\mu) D_b(x_2,\mu)+D_b(x_1,\mu) D_u(x_2,\mu)\right]d\hat{\sigma}(ub\to dth)
\end{eqnarray}
where $D_u$ and $D_b$ are the PDF of the $u$ and $b$ quark, respectively, in the protons. The colliding parton momenta in LHC laboratory frame are  
\begin{subequations}
\begin{align}
p_u&=\frac{\sqrt{s}}{2}(x_1,0,0,x_1),\\
p_b&=\frac{\sqrt{s}}{2}(x_2,0,0,-x_2),
\end{align}
\end{subequations}
in the first term of Eq.\,(\ref{eq:xsdiff}), whereas the $u$- and $b$-quark four momenta are exchanged in the second term. 
Therefore, the $b$-quark momentum is negative along the $z$-axis for half of the events and positive for the other half. In order to perform the azimuthal angle or polarization asymmetry measurements proposed in \cite{Barger:2018tqn}, we should identify the momentum of the virtual $W^+$ emitted from the $u$ (or $c$, $\bar{d}$, $\bar{s}$) quark. This is possible only when we can identify the sign of the $b$-quark momentum.

\subsection{Selecting the $b$ and $\bar{b}$ momentum direction}

Because valence quark distributions are harder than the sea quark distributions, we expect that the subprocess with negative $b$-quark momentum should have positive rapidity of the hard scattering system ($p_u+p_b=p_d+p_t+p_h$):
\begin{eqnarray}
Y(thj) =\frac{1}{2}\ln\frac{E(thj)+p_z(thj)}{E(thj)-p_z(thj)}=\frac{1}{2} \ln \frac{x_1}{x_2}.
\end{eqnarray}
 \begin{figure}[H]
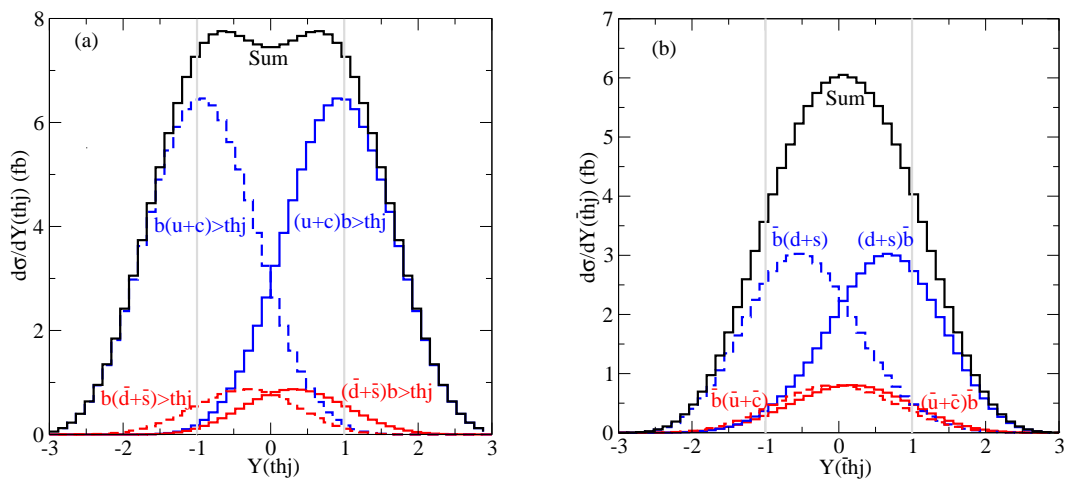

\vspace{0.5cm}
\begin{center}
\includegraphics[width=65mm]{Ythj_top.eps}\label{fig:Ytop}
\hspace{0.8cm}
\includegraphics[width=80mm]{Ythj_tbar.eps}\label{fig:Ytbar}
\caption {Rapidity distributions of the $thj$ (a) and $\bar{t}hj$ (b) systems in $pp$ collisions at $\sqrt{s}=13$ TeV. 
The blue curves are for contributions from subprocesses including the valence quark, whereas the red curves are for those from sea quark only. 
The events of $b(\bar{b})$ quark momentum direction along or opposite to the $z$ axis are shown by solid and dashed lines, respectively. 
 The black curves are the sum of the rapidities from the four subprocesses. The quark and anti-quark jets from $t$-channel W emission are tagged with cuts $p_T^j>~30~{\rm GeV}, ~{\rm and}~ |\eta_j|~<~4.5$.}
\label{fig:Ythj}
\end{center}
\end{figure}
\hspace{-0.35cm}Shown in Fig.\,\ref{fig:Ythj}(a) are the $Y$ distributions of the $thj$ events where the light quark or anti-quark jet from the $t$-channel $W$ emission are tagged with cuts.
\begin{eqnarray}
p_T^j>~30~{\rm GeV}, ~{\rm and}~ |\eta_j|~<~4.5.
\end{eqnarray}
Events with negative momentum $b$-quark are shown by solid curves, whereas those with positive $b$-quark momentum are given by dashed curves. The solid black curve shows the total sum of all $thj$ events.
The blue curves give the sum of $ub\to dth$ and $cb\to sth$ subprocess contributions (that have exactly the same matrix elements), and the red curves are for the sum of $\bar{d}b\to\bar{u}th$ and $\bar{s}b\to\bar{c}th$ subprocess contributions.
 As expected, events with $Y(thj)>1$ are mostly from the negative momentum $b$-quark (blue and red solid curves). 
 Although the purity (the probability) of negative $b$-momentum is $ 95\%$, only $41\%$ of the total events satisfy the $Y(thj)>1$ cut, leaving (59\%) of the events with mixed kinematics which results in significant reduction of observable asymmetries and polarizations.
  The situation is much worse for $\bar{t}hj$ production processes, as shown in Fig.\,\ref{fig:Ythj}(b). With the same $Y(thj)>1$ cut, the purity is only $89\%$ and only $31\%$ of the events are kept. 
  It is mainly because the down quark is not as hard and populous as the up quark in the proton. 
  All results in our study are calculated with  the CTEQ14 PDF in the LO \cite{Dulat:2015mca} with the factorization scale $\mu=\frac{m_t+m_h}{4}$, following Ref.\,\cite{Demartin:2015uha}.
\begin{figure}[t]
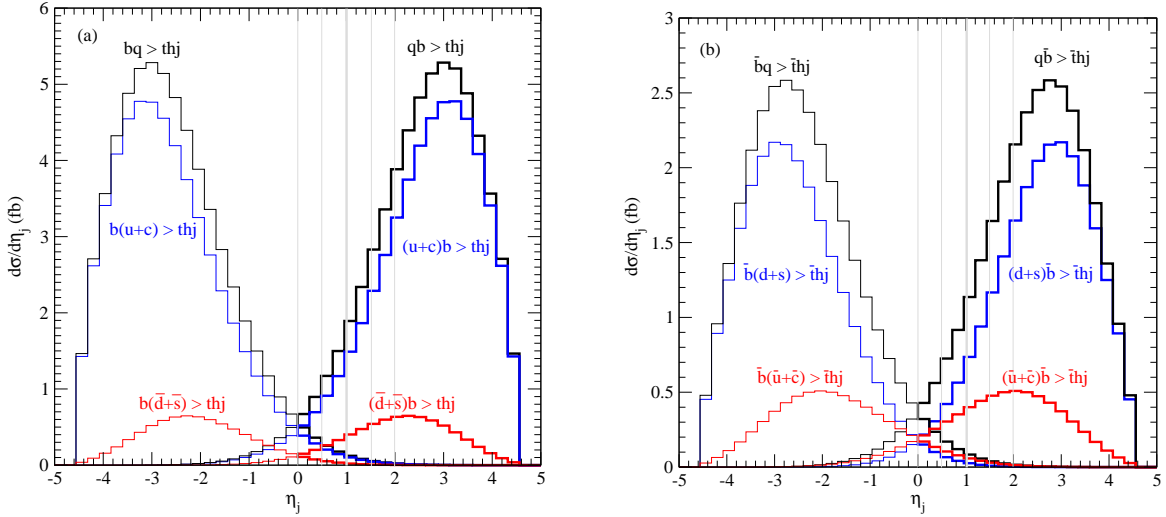

\vspace{0.5cm}
\begin{centering}
\begin{tabular}{c}
\includegraphics[width=0.4\textwidth]{Fig4_etaj.eps}\label{fig:etaj_top}
\hspace{0.8cm}
\includegraphics[width=0.4\textwidth]{etaj_tbar.eps}\label{fig:etaj_tbar}
\end{tabular}
\caption {Tagged jet pseudo-rapidity distributions in $thj$ (a) and $\bar{t}hj$ (b) events in $pp$ collisions at $\sqrt{s}=13$ TeV.}
\label{fig:etaj}
\end{centering}
\end{figure}

\begin{table}[b!]
 \begin{center}
  \begin{tabular}{| l ||ccc|ccc||c||c||cc|}
   \hline
 Cut&\multicolumn{3}{c|}{$\sigma(qb\to thj)$[fb]}&\multicolumn{3}{c||}{$\sigma(bq\to thj)$[fb]}&$\sigma(thj)$[fb]&Purity[\%]&\multicolumn{2}{c|}{Fraction in $qb$[\%]}\\
    &$ub+cb$  & $\bar{d}b+\bar{s}b$ & Sum &$bu+bc$  & $b\bar{d}+b\bar{s}$ & Sum &&$qb$&$ub+cb$  & $\bar{d}b+\bar{s}b$  \\
   \hline \hline
 $\eta_j  >0$ &  12.74 & 1.75 & 14.49&       0.32& 0.076 & 0.40 &14.89 (100\%) &97.3&87.9 & 12.1\\
 $\eta_j  >0.5$  & 12.43 & 1.66 & 14.09&    0.15& 0.031 & 0.18& 14.27 (95.8\%)&98.7 &88.2&11.8\\
  ${\eta_j  >1}$ &  11.90& 1.50& 13.40&    0.065 & 0.011 & 0.076& 13.48 (90.5\%)&99.4&88.8&11.2\\
   $\eta_j  >1.5$ & 11.02 & 1.28 & 12.30&     0.026 & 0.0033 & 0.029&  12.33 (82.8\%)&99.8&89.6&10.4\\
  $\eta_j  >2$ &  9.69 & 0.99& 10.68&    0.0093 & 0.00086 & 0.010&10.69 (71.8\%)&99.9&90.7&9.3\\  
   \hline
  \end{tabular}
  \caption{
  Cross section of $thj$ production events with cuts on
$\eta_j $.  
Contributions
of the subprocesses $(u+c)b \to thj$, $(\bar{d}+\bar{s})b \to thj$
and their sum are shown after each cut. 
 The fraction
of the events with the $b$-quark momentum along the negative $z$ direction is given as purity.
Percent values in parenthesis show the fraction
of $thj$ events which survive after the cut. 
} 
  \label{tab:dely}
 \end{center}
\end{table}
\begin{table}[t!]
 \begin{center}
  \begin{tabular}{| l ||ccc|ccc||c||c||cc|}
   \hline
 Cut&\multicolumn{3}{c|}{$\sigma(q\bar{b}\to \bar{t}hj)$[fb]}&\multicolumn{3}{c||}{$\sigma(\bar{b}q\to \bar{t}hj)$[fb]}&$\sigma(\bar{t}hj)$[fb]&Purity[\%]&\multicolumn{2}{c|}{Fraction in $q\bar{b}$[\%]}\\
    & $d\bar{b}+s\bar{b}$ & $\bar{u}\bar{b}+\bar{c}\bar{b}$  &Sum &$\bar{b}\bar{u}+\bar{b}\bar{c}$  & $\bar{b}d+\bar{b}s$ & Sum &&$q\bar{b}$  & $d\bar{b}+s\bar{b}$&$\bar{u}\bar{b}+\bar{c}\bar{b}$  \\
   \hline \hline
 $\eta_j  >0$ & 5.71&  1.47  & 7.18&        0.10 & 0.16&0.26 &7.44 (100\%) &96.5 & 79.5&20.5\\
 $\eta_j  >0.5$  & 5.58& 1.36  & 6.94&     0.038 &0.085& 0.12& 7.06 (94.9\%)&98.3 &80.4&19.6\\
  ${\eta_j  >1}$ & 5.32&  1.19  & 6.51&     0.013 &0.041 & 0.054& 6.56 (88.2\%)&99.2&81.7&18.3\\
   $\eta_j  >1.5$ & 4.88& 0.97  & 5.85&      0.0039 &0.017 & 0.021&  5.87 (78.9\%)&99.7&83.4&16.6\\
  $\eta_j  >2$ & 4.21&  0.73 &4.94 &     0.00099 &0.0065 & 0.0075&4.95 (66.5\%)&99.8&85.2&14.8\\  
   \hline
  \end{tabular}
  \caption{
 Cross section of $\bar{t}hj$ production events with cuts on $\eta_j$.  
 Contributions
of the subprocesses $(d+s)\bar{b} \to \bar{t}hj$, $(\bar{u}+\bar{c})\bar{b} \to \bar{t}hj$ and their sum are shown after each cut. 
 The fraction
of the events with the $\bar{b}$-quark momentum along the negative $z$ direction is given as purity.
Percent values in parenthesis show the fraction
of $\bar{t}hj$ events which survive after the cut. 
} 
  \label{tab:etaj_tbar}
 \end{center}
\end{table}
In Fig.\,\ref{fig:etaj}(a) and\,\ref{fig:etaj}(b), we show the tagged jet pseudo-rapidity distributions. 
Now the separation of events with negative $b$ momentum (shown by blue and red thick curves) and those with positive $b$ momentum (shown by blue and red thin curves) is clearer for both $thj$ (a) and $\bar{t}hj$ (b). 
In Tables I and II, we show the purity and the survival rate of several $\eta_j$ selection cuts for choosing events with negative $b$ or $\bar{b}$ momentum events, respectively, for $thj$ and $\bar{t}hj$ processes. 
Even for $\eta_j>0$, when all events are used in the analysis, the purity is higher than $96\%$ for both $thj$ and $\bar{t}hj$ events. In this report, we adopt the selection cut
\begin{eqnarray}
1<\eta_j<4.5,\quad\quad p_T^j>30~{\rm GeV}
\end{eqnarray}
for the jet tag. 
Since the purity is higher than $99\%$ for both $thj$ (Table \ref{tab:dely}) and $\bar{t}hj$ (Table \ref{tab:etaj_tbar}), we can safely neglect contribution from events with the wrong $b$-quark momentum direction, whose analysis requires additional kinematical considerations. 
Needless to say, events with $\eta_j<-1$ have exactly the same signal with those with $\eta_j>1$ because there is no distinction between the two colliding proton beam. 
From Tables \ref{tab:dely} and \ref{tab:etaj_tbar}, we find that the selection cut $|\eta_j|>1$ allow us to study $90\%$ of $thj$ and $88\%$ of $\bar{t}hj$ events with full kinemetical reconstruction.
In the following analysis, we assume that a significant fraction of $h$ and single $t$ or $\bar{t}$ production via $t$-channel $W$ exchange can be kinematically reconstructed, and define observables whose properties are directly determined by the helicity amplitudes of Section II. 
%
\subsection{${\tt Q}$ and ${\tt W}$ distributions}
The differential cross section for the subprocess $ub\to dth$ is 
\begin{eqnarray}
\label{eq:xsM+M-}
d\hat{\sigma}=\frac{1}{2\hat{s}}\frac{1}{4}\left(|{\cal M}_+|^2+|{\cal M}_-|^2\right)d\Phi_3(thj)
\end{eqnarray}
in terms of the helicity amplitudes ${\cal M}_\sigma$ in Eq.\,(\ref{eq:amp}), where $\hat{s}=(p_u+p_b)^2=(p_d+p_t+p_h)^2$, 1/4 is the probability to find left-handed $u$ and $b$ quarks inside their PDFs, the color factor is unity for $t$-channel color-singlet exchange between the colliding quarks, and the three-body Lorentz invariant phase space can be parametrized as 
\begin{eqnarray}
d\Phi_3(thj)=d\Phi_2(j+th)\frac{d{\tt W}^2}{2\pi}d\Phi_2(t+h) 
\end{eqnarray}
as a convolution of the two-body phase space integrated over the invariant mass ${\tt W}$ of the $t+h$ system
\begin{eqnarray}
m_t+m_h<{\tt W}<\sqrt{\hat{s}}.
\end{eqnarray}

The $j+th$ phase space
\begin{eqnarray}
d\Phi_2(j+th)=\frac{1}{8\pi}x\frac{d\cos\hat{\theta}}{2}
\end{eqnarray}
is parametrized in the $ub$ or $thj$ rest frame, where the four momenta are parametrized as 
\begin{subequations}
\begin{align}
p_u&=\frac{\sqrt{\hat{s}}}{2}(1,0,0,1),\\
 p_b&=\frac{\sqrt{\hat{s}}}{2}(1,0,0,-1)\\
  p_d&=\frac{\sqrt{\hat{s}}}{2}x(1,\sin\hat{\theta},0,\cos\hat\theta)\\
    q&=p_u-p_d=\frac{\sqrt{\hat{s}}}{2}(1-x,-x\sin\hat{\theta},0,1-x\cos\hat\theta)
\end{align}
\end{subequations}
with 
$
x=1-{\tt W}^2/\hat{s}
$,
 and 
 \begin{eqnarray}
 \label{eq:Qdefination}
 {\tt Q}^2=-q^2=\hat{s}x\frac{1-\cos\hat{\theta}}{2}.
 \end{eqnarray}
 The jet rapidity in the lab frame is hence 
\begin{eqnarray}
\eta_j=\frac{1}{2}\ln\frac{1+\cos\hat{\theta}}{1-\cos\hat{\theta}}+Y(thj).
\end{eqnarray}
The forward peak in the $\eta_j$ distribution in Fig.\,\ref{fig:etaj} is due to the square of the common $t$-channel $W$ propagator in the amplitude,
\begin{eqnarray}
|D_W(q)|^2=\frac{1}{({\tt Q}^2+m_W^2)^2}=\frac{1}{(\hat{s}x(1-\cos\hat\theta)/2+m_W^2)^2}
\end{eqnarray}
which grows towards $\cos\hat{\theta}\sim1$, subject to the jet $p_T$ cut
\begin{eqnarray}
p_T^d=\frac{\sqrt{\hat{s}}}{2}x\sin\hat{\theta} >30~{\rm GeV}
\end{eqnarray}
\begin{figure}[ b!]
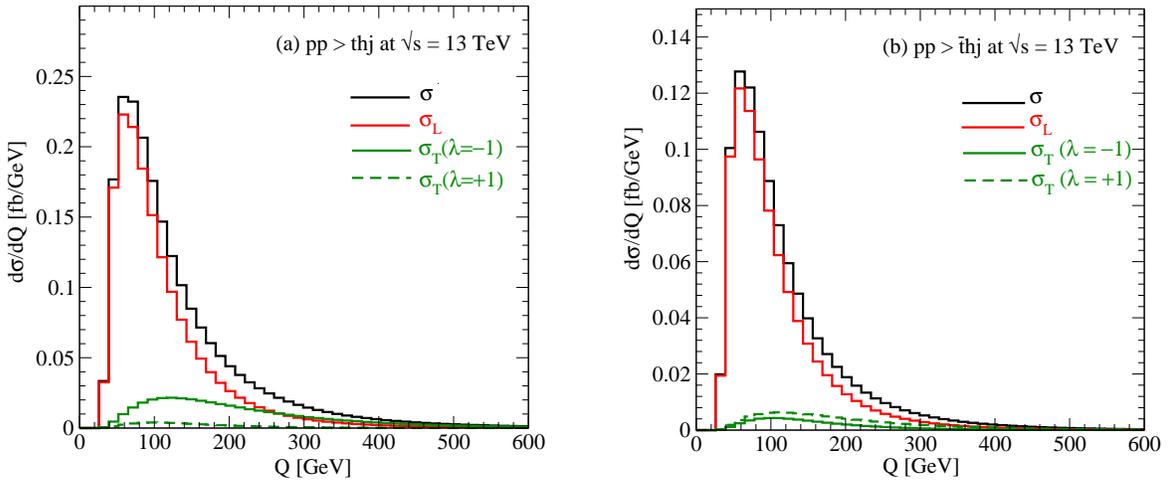

\vspace{+0.8cm}
\begin{centering}
\begin{tabular}{c}
\includegraphics[width=0.4\textwidth]{Q_top_mul2.eps}
\hspace{0.8cm}
\includegraphics[width=0.4\textwidth]{tbar_Q_mul2.eps}
\end{tabular}
\caption {$d\sigma/d{\tt Q}$ for $pp>thj$ (a) and $pp>\bar{t}hj$ in $pp$ collisions at $\sqrt{s}=13$ TeV, with the jet tag condition of $p_T^j>30~{\rm GeV}$ and $1<|\eta_j|<4.5$ . ${\tt Q}=\sqrt{-q^2}$ is the invariant momentum transfer of the virtual $W^+$ (a) or $W^-$ (b). The red curves show contributions of the longitudinal $W(\lambda=0)$, while the green curves show those of the transverse $W(\lambda=\pm1)$.}
\label{fig:Q}
\end{centering}
\end{figure}
The $t+h$ phase space in the $th$ rest frame is
\begin{eqnarray}
d\Phi_2(t+h) =\frac{1}{8\pi}\bar{\beta}\frac{d\cos\theta^\ast}{2}\frac{d\phi}{2\pi}
\end{eqnarray}
where the participating four momenta are parametrized as 

\begin{subequations}
\begin{align}
q&=\frac{\tt W}{2}\left(1-\frac{{\tt Q}^2}{\tt W^2},~0,~0,~1+\frac{{\tt Q}^2}{\tt W^2}\right)\\
p_b&=\frac{\tt W}{2}\left(1+\frac{{\tt Q}^2}{\tt W^2},~0,~0,~-1-\frac{{\tt Q}^2}{\tt W^2}\right)\\
p_t&=\frac{\tt W}{2}\left(1+\frac{m_t^2-m_h^2}{\tt W^2},~\bar{\beta}\sin\theta^\ast\cos\phi,~\bar{\beta}\sin\theta^\ast\sin\phi,~\bar{\beta}\cos\theta^\ast\right)\\
p_h&=\frac{\tt W}{2}\left(1+\frac{m_h^2-m_t^2}{\tt W^2},~-\bar{\beta}\sin\theta^\ast\cos\phi,~-\bar{\beta}\sin\theta^\ast\sin\phi,~-\bar{\beta}\cos\theta^\ast\right).
\end{align}\label{eq:qpbptph2}
\end{subequations}
\hspace{-0.1cm}When evaluating the amplitudes ${\cal M}_\sigma$, we rotate the frame about the virtual $W$ momentum axis so that the top three momentum is in the $x$-$ z$ plane, as in Eq.\,(\ref{eq:Qdefination}) and the azimuthal angle is given to the $u\to dW$ emission plane as in Eq.\,(\ref{eq:pud}) in the Breit frame.

\begin{figure}[b]
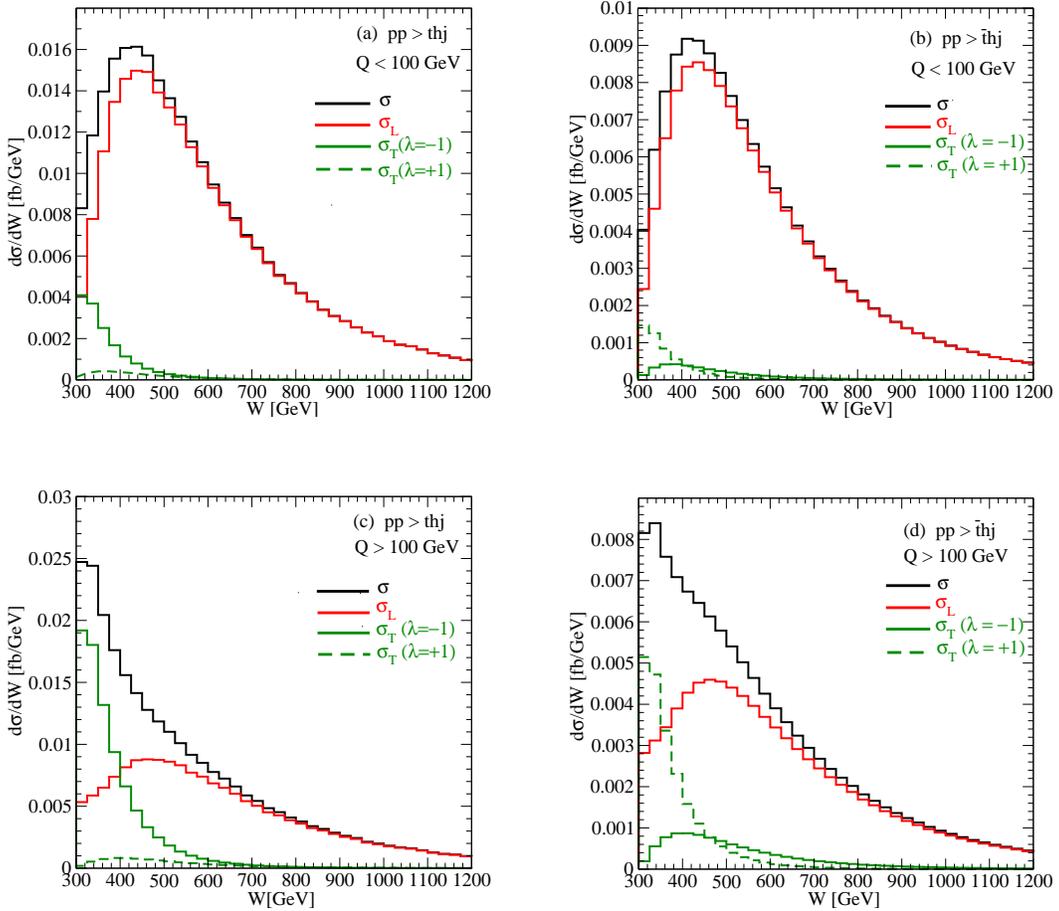

\vspace{+0.5cm}
\begin{centering}
\includegraphics[width=0.36\textwidth]{W_top_Qle100.eps}
\hspace{0.8cm}
\includegraphics[width=0.36\textwidth]{tbar_W_Qlt100.eps}\\
\vspace{+0.9cm}
\includegraphics[width=0.36\textwidth]{W_top_Qgt100.eps}
\hspace{0.8cm}
\includegraphics[width=0.36\textwidth]{tbar_W_Qgt100.eps}
\caption { $d\sigma/d{\tt W}$ v.s.\,${\tt W}$, where ${\tt W}=\sqrt{P_{th}^2}=m(th)$ is  the invariant mass of $th$ system. The upper two plots are for small ${\tt Q}$ (${\tt Q}<100$ GeV), while the lower two plots are for large ${\tt Q}$ (${\tt Q}>100$ GeV). The left plots (a) and (c) are for $thj$, while the right ones (b) and (d) are for $\bar{t}hj$. The red curves show contributions of the longitudinal $W(\lambda=0)$, while the green curves show those of the transverse $W(\lambda=\pm1)$.}
\label{fig:QW}
\end{centering}
\end{figure}
We show in Fig.\,\ref{fig:Q} the distributions with respect to the momentum transfer ${\tt Q}$, Eq.\,(\ref{eq:Qdefination}).
 Contributions from $\lambda=0$ and $\lambda=\pm1$ $W's$ separately and their sum are shown. 
 Because the momentum transfer ${\tt Q}$ does not depend on the azimuthal angle, integration over $\phi$ about the $W$-momentum axis (the common $z$-axis in Fig.\,\ref{fig:frame}) projects out the $W$ helicity states and the interference among different $\lambda$ contributions vanish.
  It is clearly seen that the longitudinal $W~(\lambda=0)$ contribution in red solid curves dominates at small ${\tt Q}~({\tt Q}\lesssim $100 GeV) both for $thj$ and $\bar{t}hj$. 
  This is a consequence of the ${\tt W}/{\tt Q}$ enhancement of the $\lambda=0$ amplitudes as shown explicitly in Eqs.\,(\ref{eq:amp}) for $thj$,  and in Eq.\,(\ref{eq:amptbar}) for $\bar{t}hj$.
   Among the transverse $W$ contributions, $\lambda=-1$ (solid green) dominates over $\lambda=+1$ (dashed green) for $thj$, but they are comparable for $\bar{t}hj$.

This somewhat different behaviour of the transverse $W$ contribution between $thj$ and $\bar{t}hj$ processes needs clarification, and we show in Fig.\,\ref{fig:QW} the distribution with respect to ${\tt W}$, the invariant mass of the $th$ system. 
The upper plots (a) and (b) are for ${\tt Q}<100$ GeV, and the lower plots (c) and (d) are for ${\tt Q}>100$ GeV. 
The left figures (a) and (c) are for $thj$, while the right ones (b) and (d) are for $\bar{t}hj$.
Again contributions from the three helicity states of the exchanged $W$ are shown separately. It is clearly seen that at small ${\tt Q}~({\tt Q}<100$ GeV) and large ${\tt W}$, ${\tt W}\gtrsim500$ GeV, the longitudinal $W~(\lambda=0)$ contribution dominates the cross sections of both $thj$ (a) and $\bar{t}hj$ (b) production. 
The transverse $W$ contributions are significant at large ${\tt Q}$ (${\tt Q}>100$ GeV), where the left handed $ (\lambda=-1)$ $W$ dominates over the longitudinal $W$ ($\lambda=0$) at ${\tt W}\lesssim400$ GeV for $thj$. 

On the other hand, the right-handed $W^-$ dominates $\bar{t}hj$ production at small ${\tt W}$, especially at large ${\tt Q}$ (${\tt Q}>100$) GeV, see Fig.\,\ref{fig:QW}(d).
This is because the $\lambda=+1$ $W^{-}$ collides with the right-handed $\bar{b}$-quark, giving $J_z=+\frac{1}{2}$ initial state with no $\bar{\beta}$ suppression, as can be seen from the first terms in Eqs.\,(\ref{eq:M+tbar}) and (\ref{eq:M-tbar}).
The $\lambda=-1$ $W^-$contribution dominates over $\lambda=+1$ at large ${\tt W}$, because the left-handed $d$-quark prefers to emit $\lambda=-1$ $W^-$, as can be seen from the $d\to uW^-$ splitting amplitudes $J_\lambda$, Eq.\,(\ref{eq:J}). 

Summing up, the $\lambda=+1~W^-$ contribution is significant near the threshold (${\tt W}\lesssim400$ GeV) for $\bar{t}hj$ production, while the $\lambda=-1$ $W^-$ contribution takes over at larger ${\tt W}$ because of dominant $d$-quark contribution. 
In contrast, for the $thj$ production, the $\lambda=+1$ contribution (green dashed lines) is deeply suppressed, as the disfavored helicity emitted from left-handed $u$ quark at large ${\tt W}$ and by the $p$-wave threshold suppression at small {\tt W}, making them very small both at small (a) and large {\tt Q} (b).


\section{azimuthal angle asymmety}\label{sec:A_azimuthal}
\begin{figure}[b!]
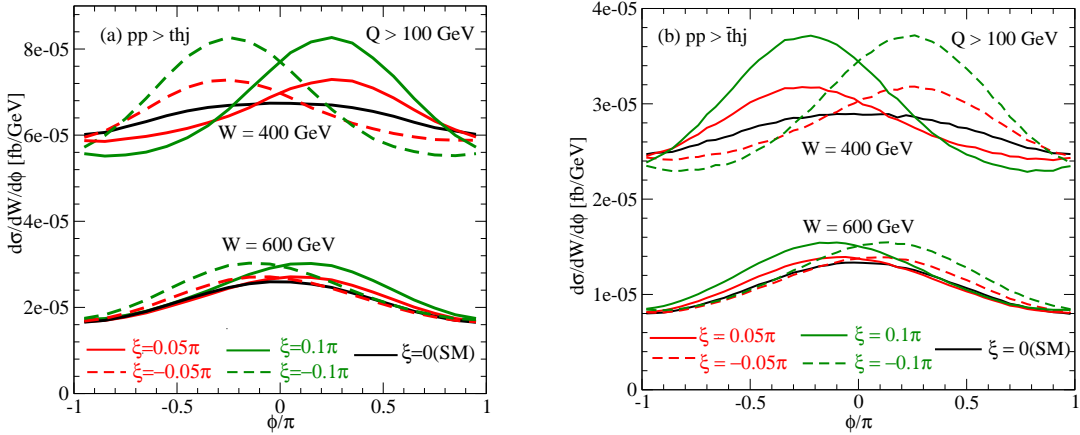

\vspace{0.5cm}
\begin{centering}
\includegraphics[width=0.36\textwidth]{phi_top.eps}\label{fig:phi_top}
\hspace{0.8cm}
\includegraphics[width=0.38\textwidth]{phi_tbar.eps}\label{fig:phi_tbar}
\caption {$d\sigma/d{\tt W}/d\phi$ v.s.\,$\phi$ at ${\tt W}=400$ and 600 GeV for ${\tt Q}>100$ GeV for $pp\to thj$ (a) and $\bar{t}hj$ (b). Black, red and green curves are for the SM ($\xi=0$), $\xi=\pm0.05\pi,~{\rm and}~\pm0.1\pi$. The solid curve are for $\xi\geq$0, while the dashed curves are for $\xi<0$.}\label{fig:phi}
\end{centering}
\end{figure}

In Fig.\,\ref{fig:phi}(a), we show distributions of the azimuthal angle between the emission ($u\to dW^+,~\bar{d}\to\bar{u}W^+,etc$) plane and the $W^+b\to th$ production plane about the common $W^+$ momentum direction in the $W^+ b$ rest frame; see Fig.\,\ref{fig:frame}. 
Shown in Fig.\,\ref{fig:phi}(b) are the same distributions for $pp\to\bar{t}hj$ process, where the azimuthal angle is between the $W^-$ emission plane and the $W^-\bar{b}\to\bar{t}h$ production plane about the common $W^-$ momentum direction.
The results are shown at ${\tt W}=400$ and 600 GeV for large ${\tt Q}$ (${\tt Q} >$ 100 GeV).
 The black, red and green curves are for the SM ($\xi=0$), $\xi=\pm0.05\pi,~{\rm and}~\pm0.1\pi$, respectively. 
 Solid curves are for $\xi\geq0$ while dashed curves are for $\xi<0$. 
The $\phi$ distributions are proportional to 
\begin{eqnarray}
|{\cal M}_+|^2+|{\cal M}_-|^2
\end{eqnarray} 
where the top polarization is summed over.Likewise, they are proportional to $|{\overline {\cal M}}_+|^2+|{\overline {\cal M}}_-|^2$ for $\bar{t}hj$ events. 
Analytic expression for the amplitudes, ${\cal M}_{\pm}$ and ${\overline{\cal M}}_\pm$ are given in Eqs.\,(\ref{eq:amp}) and (\ref{eq:amptbar}), respectively, where we can tell that azimuthal angle dependences are in the $\lambda=\pm1$ $W^\pm$ exchange amplitudes.
The asymmetry is large at small ${\tt W}$ and large ${\tt Q}$ because the transverse $W^\pm$ amplitudes are significant there, see Fig.\,\ref{fig:QW}. 
The asymmetry remains significant at ${\tt W}=400$ GeV, however, even for events with ${\tt Q}<100$ GeV~\cite{Barger:2018tqn}.

\begin{figure}[t]
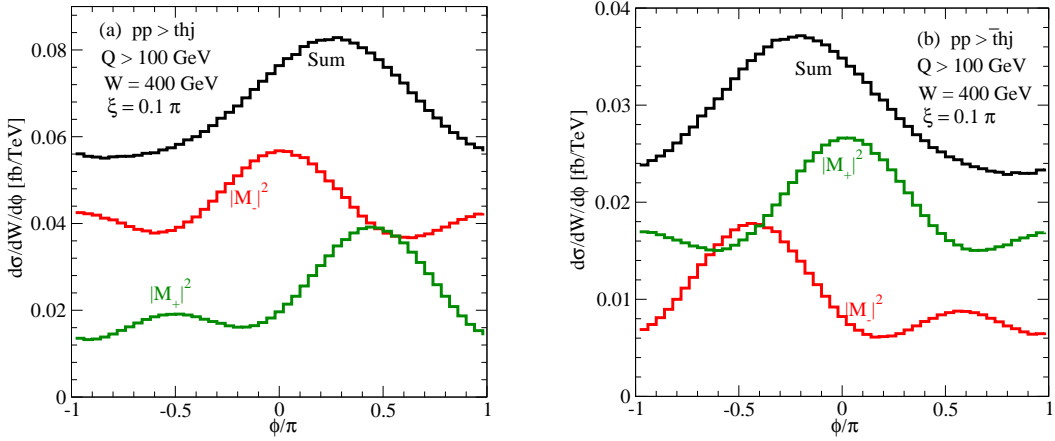

\begin{centering}
\begin{tabular}{c}
\includegraphics[width=0.36\textwidth]{phiPMQgt100_W400_shifted.eps}\label{fig:phi2}
\hspace{0.8cm}
\includegraphics[width=0.36\textwidth]{tbar_phiPMQgt100_W400_shifted.eps}\label{fig:phi2}
\end{tabular}
\caption {Azimuthal angle distribution of $|{\cal M}^2_+|$ (red) and $|{\cal M}^2_-|$ (green) and the sum (black) for $pp\to thj$ (a) and $pp\to\bar{t}hj$ (b) for ${\tt Q}>100$ GeV events at ${\tt W} =400$ GeV, for $\xi=0.1\pi$. }
\label{fig:M+-}
\end{centering}
\end{figure}

We show in Fig.\,\ref{fig:M+-}(a) the azimuthal angle distribution of right-handed and left-handed top quark separately, in green and red curves respectively, at ${\tt W}=400$ GeV for events with ${\tt Q}>100$ GeV and $\xi=0.1\pi$.
 Their sum, given by the black curve agree with the corresponding curve in Fig.\,\ref{fig:phi}(a). 
As expected from the analytic expressions Eqs.\,(\ref{eq:amp}) and (\ref{eq:simMpm}), $|{\cal M}_-|^2$ is almost symmetric about $\phi=0$, and the asymmetry is mainly from $|{\cal M_+}|^2$.
 Likewise, for the $\bar{t}hj$ events, shown in Fig.\,\ref{fig:M+-}(b), the asymmetry is mainly from left handed $\bar{t}$ quark, depicted by the red $|{\overline{\cal M}_-}|^2$ curve.

\begin{figure}[b]
\vspace{0.5cm}
\begin{centering}
\begin{tabular}{c}
\includegraphics[width=0.4\textwidth]{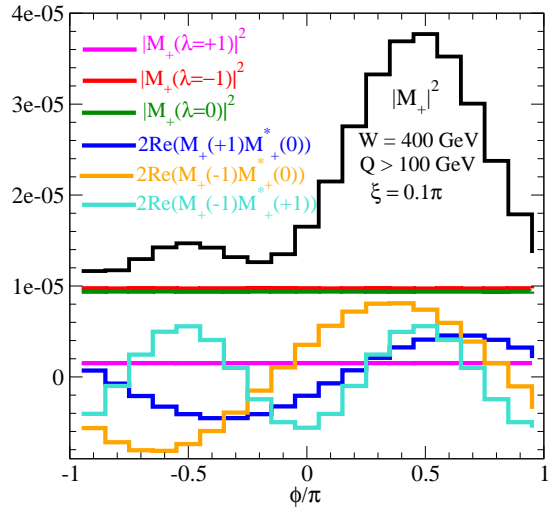}\label{fig:phi2}
\end{tabular}
\caption {Azimuthal angle distribution of $|{\cal M}_+|^2=|\sum_\lambda {\cal M}_+ (\lambda)|^2$ in terms of the six combinations of the three $W^+$ polarization amplitudes, ${\cal M}_+(\lambda)$ for $\lambda$=+1, -1, and 0. The subprocess $ub\to dth$ contribution to the $pp\to thj$ process is shown.}
\label{fig:M+123456}
\end{centering}
\end{figure}

The origin of the azimuthal angle asymmetry comes from the interference between transverse $W$ amplitudes with the $e^{\pm i\phi}$ phase factor for $\lambda=\pm1$ $W$ and the longitudinal $W$ ($\lambda=0$) amplitudes as shown in Eq.\,(\ref{eq:amp}) for $ub\to thj$ and Eq.\,(\ref{eq:amptbar}) for $d\bar{b}\to u\bar{t}h$. 
We show in Fig.\,\ref{fig:M+123456} the azimuthal angle distribution of $|{\cal M}_+|^2$ for the subprocess $ub\to dth$, together with the six individual contributions
\begin{eqnarray}
|{\cal M}_+|^2
&=&|\sum_\lambda {\cal M}_+(\lambda)|^2\nonumber\\
&=&|{\cal M}(+1)|^2+|{\cal M}(-1)|^2+|{\cal M}(0)|^2\nonumber\\
&&+2{\rm Re}{\cal M}(+1){\cal M}^\ast(0)
+2{\rm Re}{\cal M}(-1){\cal M}^\ast(0)
+2{\rm Re}{\cal M}(-1){\cal M}^\ast(+1),\,
\end{eqnarray}
\begin{figure}[t]
\begin{centering}
\begin{tabular}{c}
\includegraphics[width=0.4\textwidth]{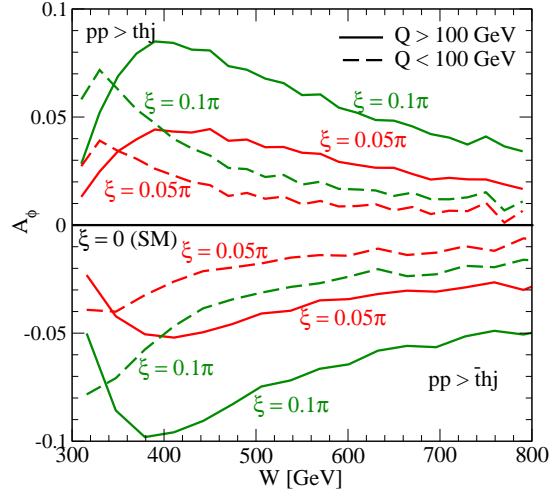}\label{fig:phi2}
\end{tabular}
\caption {%
Asymmetry $A_\phi ({\tt W})$ for $pp\to thj$ and $pp\to\bar{t}hj$ as functions of ${\tt W}$, the invariant mass of $th$ or $\bar{t}h$ system. 
Large ${\tt Q}$ (${\tt Q}>$ 100 GeV) events are shown by solid lines, while small ${\tt Q}$ (${\tt Q}<100$)GeV, events are shown by dashed curves. 
Results are shown for $\xi=0$ (SM), $\xi=0.05\pi$ (red) and $0.1\pi$ (green). $A_\phi>0$ for $th$ and $A_\phi<$0 for $\bar{t}h$, when $\xi>0$.}
\label{fig:Aphi}
\end{centering}
\end{figure}
\hspace{-0.1cm}separately. The three squared terms, $|{\cal M}(\lambda)|^2$, for $\lambda=+1,\, -1$ and 0, give no $\phi$ dependence, while the interference terms between ${\cal M}_+ (0)$ and ${\cal M}_+ (-1)$ amplitudes give terms proportional to $\sin\phi\sin\xi$ with positive coefficients, leading to positive $\langle\sin\phi\rangle$ for $\sin\xi>0$. 
It is clearly seen from Fig.\,\ref{fig:M+123456} that $|{\cal M}_+(\lambda=-1)|^2\simeq |{\cal M}_+(\lambda=0)|^2\gg|M_-(\lambda=+1)|^2$ at ${\tt W}=400$ GeV for ${\tt Q}>100$ GeV for the subprocess $ub\to dth$, consistent with the trend expected from the SM amplitudes at $\xi=0$, shown in Fig.\,\ref{fig:QW}(c). 
It is therefore the interference between the ${\cal M}_+(\lambda=-1)$ and ${\cal M}_+(\lambda=0)$ amplitudes, shown by the orange curve in Fig.\,\ref{fig:M+123456}, which determines the asymmetry $\langle\sin\phi\rangle$.
The interference between the $\lambda=\pm1$ $W$ exchange amplitudes give terms of the form $\sin2\phi\sin\xi$, which gives rise to another asymmetry $\langle\sin2\phi\rangle$.
Because $|{\cal M}_-(\lambda=+1)|$ is generally small at all ${\tt Q}$ and ${\tt W}$ regions, as shown in Figs.\,\ref{fig:QW}(a) and (c), the asymmetry $\langle\sin2\phi\rangle$ turns out to be small in our analysis. We therefore do not show results on $\langle\sin2\phi\rangle$ in the following, but note that its measurement should improve the $\xi$ sensitivity at a quantitive level, and that it should be sensitive to other type of new physics that affects mainly the transversally polarized $W$ amplitudes. It may be worth noting that asymmetry $\langle\sin2\phi\rangle$ is larger in $\bar{t}hj$ process, because both $\lambda=\pm1$ transversally polarized $W$ contributions are significant, as can be seen from Fig.\,\ref{fig:QW}(d), especially at large ${\tt Q}$ and small ${\tt W}$.

\begin{figure}[b]
\vspace{0.5cm}
\begin{centering}
\begin{tabular}{c}
\includegraphics[width=0.4\textwidth]{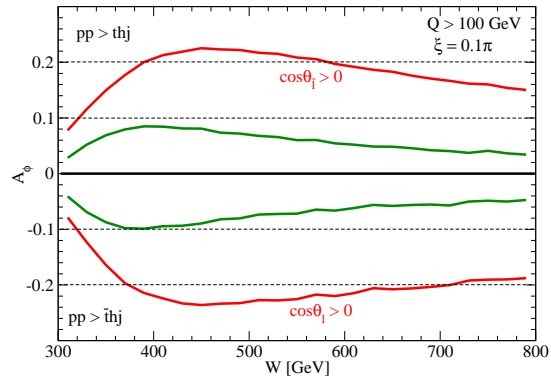}\label{fig:phi2}
\end{tabular}
\caption {
Asymmetry $A_\phi({\tt W})$ as a function of invariant mass ${\tt W}$ of the $th$ or $\bar{t}h$ system in $pp\to thj$, or $\bar{t}hj$ process, when the momentum transfer ${\tt Q}$ of the exchanged ${ W}$ is larger than 100 GeV. 
$A_\phi({\tt W})=0$ for the SM $(\xi=0)$. 
$A_\phi ({\tt W})>0$ for $thj$, while $A_\phi ({\tt W})<0$ for $\bar{t}hj$ when $\xi=0.1\pi$. The green solid curves are the asymmetries without cuts. 
The asymmetry grows to the red curves when the charged lepton decay angle satisfy $\cos\theta_{\bar{\ell}},\cos\theta_{\ell}>0$ in the $t$ and $\bar{t}$ rest frame.
}
\label{fig:Aphicheck}
\end{centering}
\end{figure}

In Fig.\,\ref{fig:Aphi}, we show the azimuthal asymmetry integrated over $\phi$,
\begin{eqnarray}
A_\phi({\tt W})=\frac{\int_{-\pi}^\pi d\phi~ {\rm sgn}(\phi) d\sigma/d{\tt W}/d\phi}{d\sigma/d{\tt W}}
\end{eqnarray}
as a function of the invariant mass ${\tt W}$ of the $th$ or $\bar{t}h$ system for $\xi=0$ (SM), $\pm 0.05\pi$ (red curve) and $\pm0.1\pi$ (green curve). 
The asymmetry for large ${\tt Q}$ (${\tt Q}>100$) GeV events is shown by solid curves, while those for small ${\tt Q}$ (${\tt Q}<100$ GeV) is shown by dashed curves. 
The positive aysmmetry is found for $thj$ events, while negative asymmetry is found for $\bar{t}hj$, in accordance with the observation from the $\phi$ distribution in Fig.\,\ref{fig:phi}. 
Generally speaking, the asymmetry is large for large ${\tt Q}$ events at around ${\tt W}\sim400$ GeV where the magnitudes of the transverse and longitudinal $W$ exchange amplitudes are comparable in Fig.\,\ref{fig:QW}(c) and (d). 
For small ${\tt Q}$, (${\tt Q}<100$ GeV), the asymmetry is significant only near the threshold, ${\tt W}\sim m_t+m_h$, where the transverse $W$ amplitudes are non-negligible in Fig.\,\ref{fig:QW}(a) and (b).

Because the asymmetry due to the term linear in $\sin\xi$ are nearly absent in $|{\cal M}_-|^2$ for $thj$ and in $|{\overline {\cal M}}_+|^2$ for $\bar{t}hj$, as can be seen from Eqs.\,(\ref{eq:M-}) and (\ref{eq:M+tbar}) for $\delta\simeq\delta^\prime$ approximation, we can expect enhancement of the asymmetry by selecting right-handed top and the left-handed anti-top. 
This can easily be achieved when $t$ and $\bar{t}$ decay semileptonically, where the charged-lepton decay angular distribution in the $t$ or $\bar{t}$ rest frame takes the form\,\cite{Atwood:1992vj}
\begin{subequations}
\begin{align}
&\frac{d\Gamma(t\to b\bar{\ell}\nu)}{d\cos\theta_{\bar{\ell}}}\sim(1+\sigma\cos\theta_{\bar{\ell}})^2,\\
&\frac{d\Gamma(\bar{t}\to \bar{b}\ell\bar{\nu})}{d\cos\theta_{{\ell}}}\sim(1-\bar{\sigma}\cos\theta_{{\ell}})^2,
\end{align}\label{eq:subtop}
\end{subequations}
\hspace{-0.12cm}about the helicity axis, where $\sigma$ and $\bar{\sigma}$ are twice the helicities of $t$ and $\bar{t}$, respectively, in the $th$ or $\bar{t}h$ rest frame.
 For instance, if we select those events with 
\begin{eqnarray}\label{eq:cosgt0}
\cos\theta_{\bar{\ell}},~\cos\theta_{{\ell}}>0,
\end{eqnarray}
then $d\sigma/ d{\tt W}/d\phi$ is proportional to 
\begin{subequations}
\begin{align}
&\frac{7}{8}|{\cal M}_+|^2+\frac{1}{8}|{\cal M}_-|^2\quad {\rm for~ top},\\
&\frac{1}{8}|{\overline {\cal M}}_+|^2+\frac{7}{8}|{\overline{\cal M}}_-|^2\quad {\rm for~ anti}\text{-}{\rm top},
\end{align}
\end{subequations}
and the asymmetry is significantly larger, as shown in Fig.\,\ref{fig:Aphicheck} for $\xi=0.1\pi$ when ${\tt Q}>100$ GeV.
The asymmetries shown by the green curves are when no cuts are applied, and they agree with the corresponding curves in Fig.\,\ref{fig:Aphi}.
The asymmetry grows to $A_\phi ({\tt W})\sim0.22$ for $thj$ events and $A_\phi({\tt W})\sim-0.23$ for $\bar{t}hj$ events, both at around ${\tt W}\sim450$ GeV with the selection cut of the $t$ and $\bar{t}$ decay charged lepton angles in Eq.\,(\ref{eq:cosgt0}).
%


\section{Polarization asymmetries}\label{sec:pol}

We are now ready to discuss the polarization of the top quark in the single top$+h$ production processes. 
We first note that the helicity amplitudes ${\cal M}_+$ and ${\cal M}_-$ in Eq.\,(\ref{eq:amp}) for the subprocess $ub\to dth$, and those in Eq.\,(\ref{eq:amptbar}) for $d\bar{b}\to u\bar{t}h$ are purely complex numbers when production kinematics ($\sqrt{\hat{s}},~{\tt Q},~{\tt W},\cos\tilde{\theta},~\cos\theta^\ast,~\phi$) are fixed. 
This is a peculiar feature of the SM where only the left-handed $u,~d,~{\rm and}~b$ quarks, and their anti-particles with right-handed helicities contribute to the single $t$ and $\bar{t}$ production process via $W$ exchange. 
It implies that the produced top quark polarization state is expressed as the superposition
\begin{eqnarray}\label{eq:topqm}
\left|t\right\rangle=\frac{{\cal M}_+^{}\left|J_z=+\frac{1}{2}\right\rangle+{\cal M}_-^{}\left|J_z=-\frac{1}{2}\right\rangle}{\sqrt{|{\cal M}_+^{}|^2+|{\cal M}_-^{}|^2}}
\end{eqnarray}
in the top quark rest frame, where the quantization axis is along the top momentum direction in the $th$ rest frame, where the top quark helicity is defined. The top quark is hence in the pure quantum state with $100\%$ polarization, with its orientation fixed by the complex number ${\cal M}_-/{\cal M}_+$.
 Its magnitude $|{\cal M}_-/{\cal M}_+|$ determines the polar angle and the phase ${\rm arg}({\cal M}_-/{\cal M}_+)$ determines the azimuthal angle of the top spin direction
 \footnote{See Appendix\,\ref{sec:AppA} for a pedagogical review of quantum mechanics.}. 
  Therefore, the kinematics dependence of the polarization direction can be exploited to measure the CP phase $\xi$, e.g.\ by combining matrix element methods with the polarized top decay density matrix
  \footnote{The top quark decay polarization density matrices for its semi-leptonic and hadronic decays are given in Appendix\,\ref{sec:AppB}.}. 
  Exactly the same applies for the $\bar{t}$ spin polarization, whose quantum state can be expressed as in Eq.\,(\ref{eq:topqm}) where the helicity amplitudes ${\cal M}_{\pm}$ are replaced by $\overline {{\cal M}}_\pm$.

In this report, we investigate the prospects of studying CP violation in the $htt$ coupling through the top and anti-top quark polarization asymmetries in the single $th$ and $\bar{t}h$ processes respectively, with partial integration over the final state phase space. For this purpose, we introduce a complex matrix distribution
\begin{eqnarray}
\label{eq:dsigMatrix}
d\sigma_{\lambda\lambda^\prime}^{}
=
\int dx_1\int dx_2 
\left[
D_{u}(x_1,\mu) D_{b}(x_2,\mu)
+
D_{b}(x_1,\mu) D_{u}(x_2,\mu)
\right]
\frac{1}{2\hat{s}}
 \frac{1}{4} M_\lambda M_{\lambda^\prime}^\ast d\Phi_{3}(dth)
\end{eqnarray}
Note that the matrix (\ref{eq:dsigMatrix}) is normalized such that its trace gives the differential cross section of Eq.\,(\ref{eq:xsdiff}).
\begin{eqnarray}
d\sigma=d\sigma_{++}+d\sigma_{--}
\end{eqnarray}
Here we denote $\lambda/2$ and $\lambda^\prime/2$ for the top helicity, and the $1/4$ factor accounts for the colliding parton spin average, just as in Eq.\,(\ref{eq:xsM+M-}) for the subprocess cross section. 
All the other subprocesses which contribute to the same $thj$ final state, $cb\to sth$, $\bar{d}b\to \bar{u}th$, $\bar{s}b\to \bar{c}th$, whose matrix elements are given in Eqs.\,(\ref{eq:Mub=cb}) and (\ref{eq:Mdbarb=sbarb}) should be summed over in the matrix (\ref{eq:dsigMatrix}).
The integration over phase space and the summation over different subprocess contribution make the top quark in the mixed state and its polarization density matrix is given by
\begin{eqnarray}\label{eq:rho}
\rho_{\lambda\lambda^\prime}^{}
=\frac{d\sigma_{\lambda\lambda^\prime}^{}}{d\sigma_{++}+d\sigma_{--}}
=\frac{1}{2}\left[\delta_{\lambda\lambda^\prime}+\sum_{k=1}^{3}P_k\sigma_{\lambda\lambda^\prime}^k\right]
\end{eqnarray}
for an arbitrary distribution. The coefficients of the three $\sigma$ matrices makes a three-vector, $\vec{P}=(P_1,P_2,P_3)$, whose magnitude $P=|\vec{P}|$ gives the degree of polarization ($P=1$ for $100\%$ polarization, $P=0$ for no polarization), while its spatial orientation gives the direction of the top quark spin in the top rest frame. 
The polarization vector $\vec{P}$ in (\ref{eq:rho}) can be obtained directly from the matrix distribution (\ref{eq:dsigMatrix}) as follows
 \begin{subequations}
 \label{eq:P123}
\begin{align}
P_1&=\frac{2{\rm Re}\int d\sigma_{+-}}{\int d\sigma_{++}+\int d\sigma_{--}},\\
P_2&=\frac{-2{\rm Im}\int d\sigma_{+-}}{\int d\sigma_{++}+\int d\sigma_{--}},\label{eq:P2}\\
P_3&=\frac{\int d\sigma_{++}-\int d\sigma_{--}}{\int d\sigma_{++}+\int d\sigma_{--}},
\end{align}
 \end{subequations}
where the integral over the phase space can be chosen appropriately in order to avoid possible cancellation of polarization asymmetries. 
For the helicity amplitudes (\ref{eq:amp}) calculated in the $th$ rest frame, the $z$-axis is along the top momentum in the $th$ rest frame, and the $y$-axis is along the $\vec{q}\times\vec{p}_t$ direction, perpendicular to the $W^+ b\to th$ scattering plane. 
In Appendix A, we obtain the orientation of the top quark spin in terms of the helicity amplitudes for a pure state and for general mixed states. 

The polarization of $\bar{t}$ quark is obtained also from the matrix distribution (\ref{eq:dsigMatrix}) with the $\bar{t}hj$ amplitudes ${\overline{\cal M}}_{\bar{\lambda}}{\overline{\cal M}}_{\bar{\lambda}^\prime}^\ast$, simply by replacing $\lambda\lambda^\prime$ by $\bar{\lambda}\bar{\lambda}^\prime$ in the density matrix (\ref{eq:rho}). The orientation of the polarization vector is measured in the same frame, where the $z$-axis is now along the $\bar{t}$ quark momentum direction in the $\bar{t}h$ rest frame and the $y$-axis is along the $\vec{q}\times\vec{p}_{\bar{t}}$ direction. 

\begin{figure}
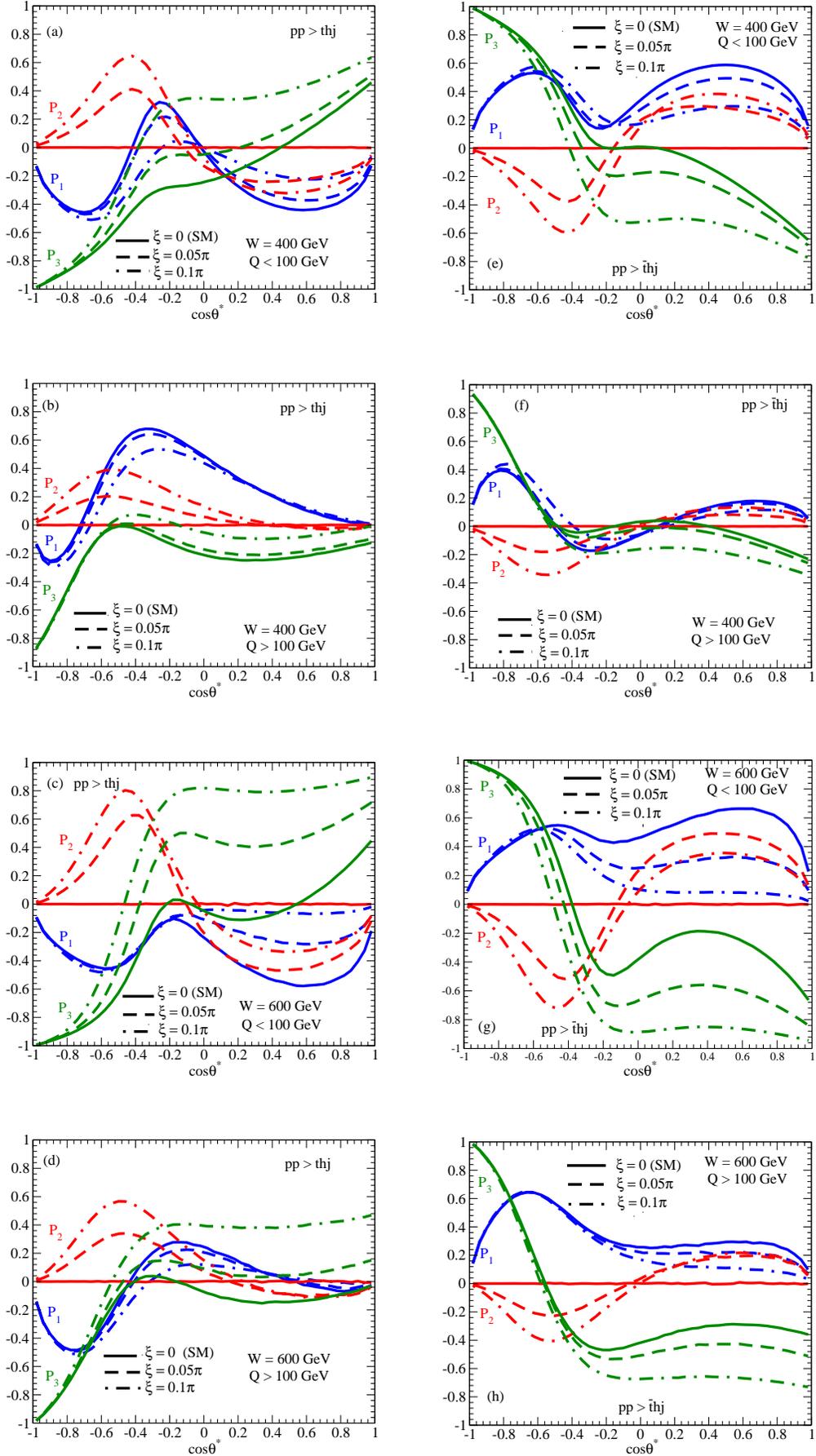

\begin{center}
\includegraphics[width=0.34\textwidth]{P_W400_Qlt100.eps}
\hspace{0.8cm}
\includegraphics[width=0.34\textwidth]{P_W400_Qlt100_tbar.eps}\\
\vspace{0.9cm}
\includegraphics[width=0.34\textwidth]{P_W400_Qgt100.eps}
\hspace{0.8cm}
\includegraphics[width=0.34\textwidth]{P_W400_Qgt100_tbar.eps}\\
\vspace{0.9cm}
\includegraphics[width=0.34\textwidth]{P_W600_Qlt100.eps}
\hspace{0.8cm}
\includegraphics[width=0.34\textwidth]{P_W600_Qlt100_tbar.eps}\\
\vspace{0.9cm}
\includegraphics[width=0.34\textwidth]{P_W600_Qgt100.eps}
\hspace{0.8cm}
\includegraphics[width=0.34\textwidth]{P_W600_Qgt100_tbar.eps}
\caption {
$t$ and $\bar{t}$ quark polarization vector $\vec{P}=( P_1,P_2,P_3)$ as function of $\cos\theta^\ast$, the $t$ or $\bar{t}$ scattering angle in the $W^+b\to th$ ($W^-\bar{b}\to\bar{t}h$) scattering plane at ${\tt W}=400$ GeV, ${\tt W}=600$ GeV for ${\tt Q}>100$ GeV and ${\tt Q}<100$ GeV. 
The LO predictions are shown for the SM ($\xi=0$) by solid curves, $\xi=0.05\pi$ by dashed curves and for $\xi=0.1\pi$ by dash-dotted curves. 
 The left four panels (a) to (d) are for $pp\to thj$ production and the right four panels (e) to (h) are for $pp\to\bar{t}hj$ production. 
 $P_2$ (denoted by the red curves) is the polarization component perpendicular to the scattering plane. 
 $P_2=0$ in the SM ($\xi=0$).}
\label{fig:pol4}
\end{center}
\end{figure}

We show in Fig.\,\ref{fig:pol4} the three components ($P_1,~P_2,~P_3$) of the polarization vector $\vec{P}$ as a function of the top (anti-top) scattering angle $\cos\theta^\ast$ in the $th$ ($\bar{t}h$) rest frame, at ${\tt W}=400$ GeV (upper four plots) and 600 GeV (lower four plots),
 when all the other kinematical variables are integrated over subject to the constraint ${\tt Q}<100$ GeV (a), (e), (c), (g) and ${\tt Q}>100$ GeV (b), (f), (d), (h).
 The left-hand side of Fig.\,\ref{fig:pol4} gives the top polarization in $thj$ processes, while the right-hand side plots give the $\bar{t}$ polarization in $\bar{t}hj$ processes.

 Let us first examine the top polarization in Fig.\,\ref{fig:pol4} (a), (b), (c), (d). 
 It is the polarization perpendicular to the scattering plane, 
 \begin{eqnarray}
 \label{eq:P2_3product}
 P_2=\langle\frac{\vec{q}\times\vec{p}_t\cdot\vec{P}}{|\vec{q}\times\vec{p}_t|}\rangle
 =\langle\frac{\vec{p}_b\times\vec{p}_h\cdot\vec{P}}{|\vec{p}_b\times\vec{p}_h|}\rangle
 \end{eqnarray}
 which vanishes for $\xi=0$ in the SM in the tree-level, as shown by red solid curves in all the plots.
 $P_1$, $P_2$ and $P_3$ are given by blue, red and green curves, respectively. 
 The three curves for each polarization components, $P_k$ ($k=1,\,2,\,3$), are for $\xi=0$ (solid curves), $\xi=0.05\pi$ (dashed curves) and for $\xi=0.1\pi$ (dot-dashed curve). 
  Significant $P_2$ polarization is expected even for $\xi=0.05\pi$ shown by red-dashed curves. 
  The plots on the right side (e), (f), (g), (h) give $\bar{t}$ polarization in the anti-top and $h$ production events. 
 In (a), (b), (c), (d), $P_3\approx-1$ (and hence $P_1\approx P_2\approx0$) at $\cos\theta^\ast=-1$, because $|{\cal M}_-|\gg|{\cal M}_+|$ at $\cos\frac{\theta^\ast}{2}=0$ and $\sin\frac{\theta^\ast}{2}=1$ in Eq.\,(\ref{eq:amp}) for $ub\to dth$ and $cb\to sth$, and similarly in Eq.\,(\ref{eq:Mdbarb=sbarb}) for $\bar{d}b\to\bar{u}th$ and $\bar{s}b\to\bar{c}th$, which have the same $W^+b\to th$ amplitudes.
 The $\lambda=\pm1$ amplitudes in ${\cal M}_+$ are strongly suppressed at $\cos\theta^\ast=-1$, while the $\lambda=0$ amplitudes in ${\cal M}_-$ are not only non-vanishing at $\cos\theta^\ast=-1$ but also ${\tt W}/{\tt Q}$ enhanced. 
 The magnitude of ${\cal M}_+$ grows quickly as $\cos\theta^\ast$ deviates from -1, and the interference between ${\cal M}_+$ and ${\cal M}_-$ gives nontrivial polarization of the top quark. 
 
 As $\cos\theta^\ast$ deviates from $-1$, $P_3$ deviates from $-1$ according to the growth of $|{\cal M}_+|^2/|{{\cal M}_-}|^2$, but $|P_1|$ (and also $|P_2|$ when $\xi\neq0$) grows quickly as they are linear in ${\cal M}_+$. 
 The polarization $P_2$ normal to the scattering plane can become as large as 0.6 even for $\xi=0.05\pi$, when ${\tt Q}<100$ GeV at ${\tt W}=600$ GeV; see Fig.\,\ref{fig:pol4}(c). 
 This is because at small ${\tt Q}$ and large ${{\tt W}}$, the longitudinal $(\lambda=0)$ $W$ contribution dominates over the transverse $(\lambda=\pm1)$ $W$ contributions, and hence the integration over the azimuthal angle $\phi$ does not diminish much the degree of top polarization. 
 
 Likewise, the $\bar{t}$ polarization is shown in the right hand side of Fig.\,\ref{fig:pol4}, for the same configuration of ${\tt W}=400$ GeV $(e), (f)$ and 600 GeV $(g), (h)$, for ${\tt Q}<100$ GeV $(e),\,( g)$ and ${\tt Q}>100$ GeV $(f), (h)$. 
 $P_3$ is now almost unity at $\cos\theta^\ast=-1$, because $|\overline {\cal M}_-(\theta^\ast)|=|{\cal M}_+(\theta^\ast)|\approx0$ at $\theta^\ast=\pi$. 
 As $\cos\theta^\ast$ deviates from $-1$, $P_3$ decreases rapidly and the polarization perpendicular to the helicity axis, $P_1$ inside the scattering plane and $P_2$ normal to the scattering plane when $\xi\neq0$ grows, just as in the case of top polarizations shown in the left-hand plots of the figure. 
 Most notably, the magnitude of all three polarization components $P_1,\, P_2,\, P_3$ behave very similar as functions of $\cos\theta^\ast$ between the top and the anti-top polarizations for the same CP phase, whereas their signs are all opposite. 
 As for $P_2$, the magnitude becomes the largest for ${\tt Q}<100$ GeV events at ${\tt W}=600$ GeV, as shown in Fig.\,\ref{fig:pol4}(c) for top and (g) for anti-top. 
 As we will explain carefully in the next section, this is a consequence of CP violation in CPT invariant theory in the absence of rescattering phase in the amplitudes. 
 
\begin{figure}[t]
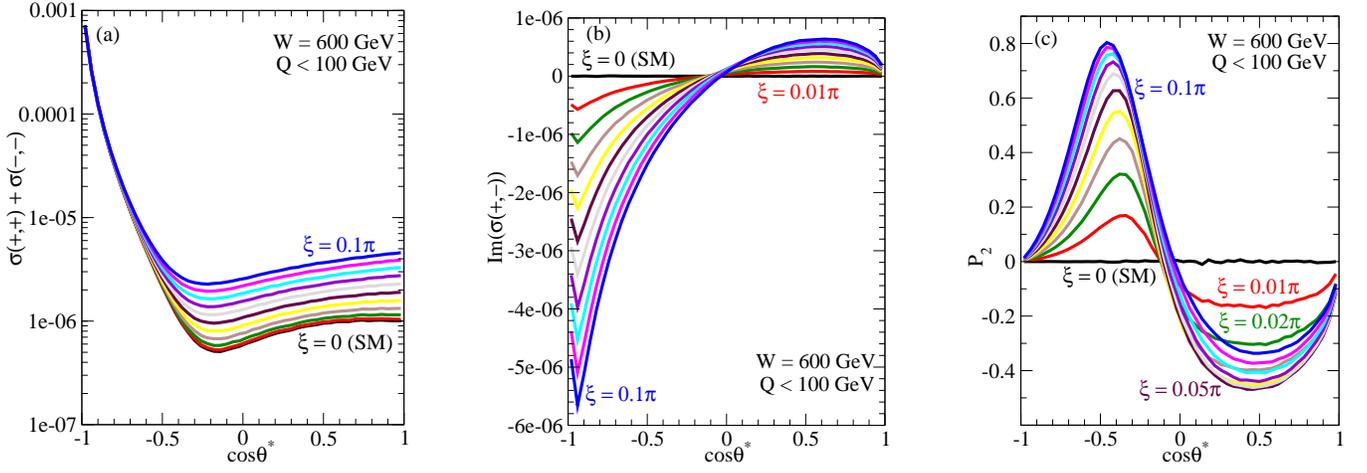

\begin{centering}
\includegraphics[width=0.3\textwidth]{intpm.eps}
\hspace{0.8cm}
\includegraphics[width=0.3\textwidth]{Im.eps}
\hspace{0.8cm}
\includegraphics[width=0.28\textwidth]{P2structure.eps}
\end{centering}
\caption{ 
Details of the $\xi$-dependence of the top quark helicity matrix distributions and $P_2$ in $pp\to thj$ at ${\tt W}=600$ GeV and ${\tt Q}<100$ GeV.
(a) $d\sigma/d\cos\theta^\ast=\sigma(+,+)+\sigma(-,-)$ is plotted against $\cos\theta^\ast$, the $t$-quark scattering angle in the $W^+b\to th$ rest frame. 
The 11 curves are for $\xi=0$ (SM) and $\xi=n\cdot 0.01\pi$ with $n=1$ to 10.
(b) Imaginary part of the off-diagonal element of the matrix distribution $d\sigma_{\lambda\lambda^\prime}$.
(c) The $t$ quark polarization perpendicular to the $W^+b\to th$ scattering plane, $P_2=-2{\rm Im}\sigma(+,-)/(\sigma(+,+)+\sigma(-,-))$. 
}
\label{fig:structure}
\end{figure}

\begin{figure}[b!]
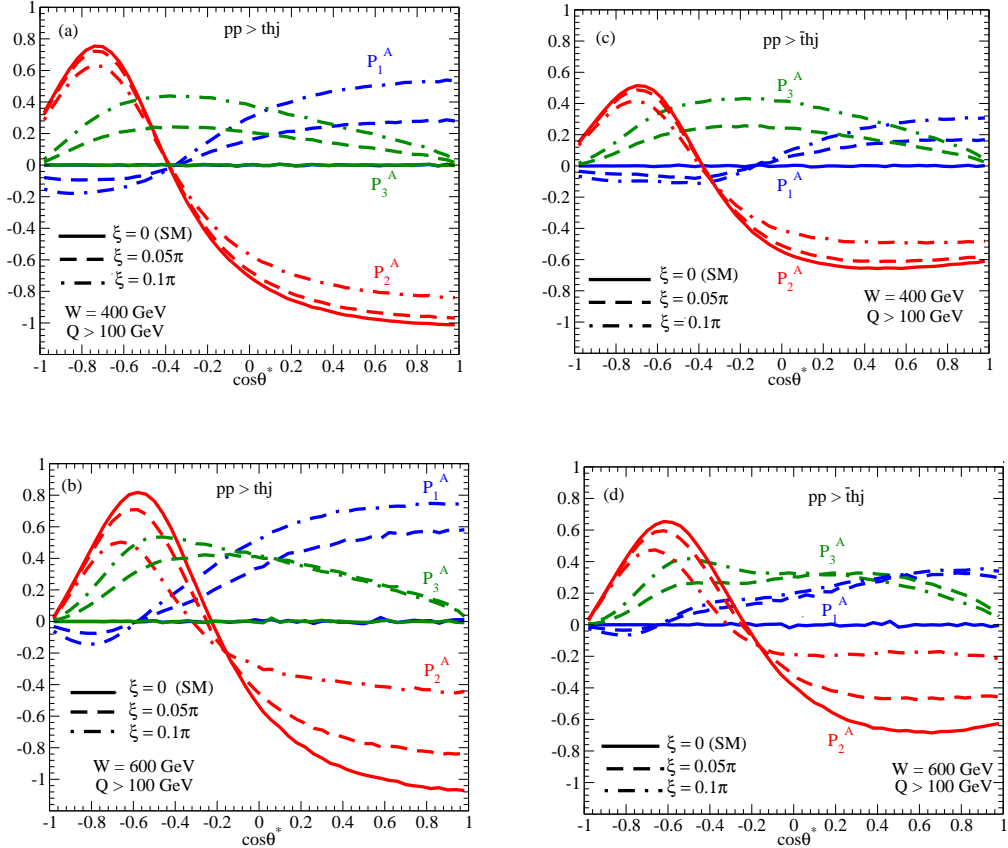

\vspace{0.5cm}
\begin{centering}
\includegraphics[width=0.34\textwidth]{PA_W400_Qgt100.eps}
\hspace{0.8cm}
\includegraphics[width=0.36\textwidth]{PA_W400_Qgt100_tbar.eps}\\
\vspace{0.9cm}
\includegraphics[width=0.34\textwidth]{PA_W600_Qgt100.eps}
\hspace{0.8cm}
\includegraphics[width=0.34\textwidth]{PA_W600_Qgt100_tbar.eps}
\end{centering}
\caption{ $P_{1,2,3}^A$ v.s.\,$\cos\theta^\ast$ for ${\tt Q}>100$ GeV events at $-1$ for ${\tt W}=400$ $(a),~(c)$, and 600 GeV $(b),~(d)$. 
The left-hand plots $(a),\,(b)$ are for $pp\to thj$ events, while the right-hand plots $(c),~(d)$ are for $pp\to\bar{t}hj$ events. 
 The solid, dashed, and dash-dotted curves are for $\xi=0$ (SM), $\xi=0.05\pi$ and $\xi=0.1\pi$, respectively. 
 $P_1^A=P_3^A=0$ for $\xi=0$ (SM).
 }
 \label{fig:pol5}
\end{figure}

Before we move on studying $t$ and $\bar{t}$ polarization after integration over $\cos\theta^\ast$, we note in Fig.\,\ref{fig:pol4}(c) and (g) for ${\tt Q}<100$ GeV at ${\tt W}=600$ GeV, the magnitudes of $P_2$ are predicted to be larger for $\xi=0.05\pi$ (dashed red curve) than those for $\xi=0.1\pi$ (dash-dotted curve) in the $\cos\theta^\ast>0$ region. 
This non-linear behavior was not expected for relatively small phase of $|\xi|\leq0.1\pi$, and we study the elements of matrix $d\sigma_{\lambda\lambda^\prime}$ carefully for ten values of $\xi$ in the range $0<\xi<0.1\pi$. 
Shown in the left plot of Fig.\,\ref{fig:pol5} is the $thj$ production differential cross section, $\sigma_{++}+\sigma_{--}$, with respect to $\cos\theta^\ast$ at ${\tt W}=600$ GeV for ${\tt Q}<100$ GeV events.
 The cross section is smallest at $\xi=0$, and grows with $\xi$ almost linearly in the region $\cos\theta^\ast\gtrsim-0.5$. 
The cross section near $\cos\theta^\ast=-1$ is dominated by the $W$ exchange amplitudes (with the $A$ factor), and hence does not depend on the $htt$ coupling.
 In the middle plot, Fig.\,\ref{fig:structure}(b), we show ${\rm Im}(\sigma(+,-))$ v.s.\,$\cos\theta^\ast$.
  Its magnitude grows with $\xi$, but it changes sign at around $\cos\theta^\ast=0$ and the growth of the magnitude is very slow at $\cos\theta^\ast>0$. 
  The average polarization $P_2$ is obtained as their ratio $-2{\rm Im}\sigma(+,-)/(\sigma(+,+)+\sigma(-,-))$ in Eq.\,(\ref{eq:P2}), which is shown in Fig.\,\ref{fig:structure}(c). 
  In the $\cos\theta^\ast>0$ region, the magnitude grows from $\xi=0$ up to $\xi\simeq0.05\pi$, but decreases to the orange curve at $\xi=0.1\pi$. 
  This study shows that the polarization $P_2$ has strong sensitivity to the CP phase $\xi$, whose magnitude can reach $20\%$ even for $\xi=\pm0.01\pi$. 

\begin{figure}[h]
\vspace{0.5cm}
\begin{centering}
\includegraphics[width=0.4\textwidth]{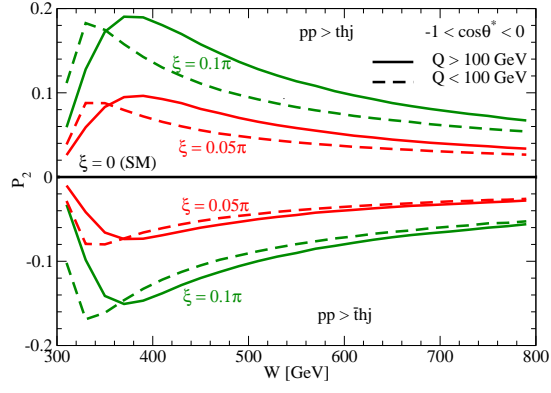}
\caption { 
$P_2$ v.s.\,${\tt W}$ for $pp\to thj$ (a) and $pp\to\bar{t}hj$ (b) in the region $-1<\cos\theta^\ast<0$. 
The green curves are for $\xi=0.1\pi$, while the red curves are for $\xi=0.05\pi$. 
The soid curves are for ${\tt Q}>$ 100 GeV, while the dashed curves are for ${\tt Q}<100$ GeV.}
\label{fig:P2_W}
\end{centering}
\end{figure}

As can be seen from Fig.\,\ref{fig:structure}(a), the differential cross section decreases sharply as $\cos\theta^\ast$ deviates from $\cos\theta^\ast=-1$, and hence the polarization asymmetry integrated over $\cos\theta^\ast$ is determined by the sign and magnitude in the $\cos\theta^\ast<0$ region.
Shown in Fig.\,\ref{fig:P2_W} are the polarization asymmetry $P_2$ for top (above zero) and antitop (below zero), for the events with $\cos\theta^\ast<0$, plotted against the $th~ (\bar{t}h)$ invariant mass ${\tt W}$. 
The results for ${\tt Q}>100$ GeV are shown by solid curves, while those for ${\tt Q}<100$ GeV are shown by dashed curves. 
The red curves are for $\xi=0.05\pi$, while green curves are for $\xi=0.1\pi$. 
Although the ad-hoc selection cut $\cos\theta^\ast<0$ is not optimal, we can observe the general trend that the magnitude of the polarization asymmetry $P_2$ grows with the CP phase $\xi$, and the sign of $P_2$ is positive for $t$, but it is negative for $\bar{t}$, when $\xi>0$. 

We may tempt to conclude that the same physics governs the sign of $A_\phi$ in Fig.\,\ref{fig:Aphi} and that of $P_2$ in Fig.\,\ref{fig:P2_W}, since both asymmetries change sign between $thj$ and $\bar{t}hj$ events. We will study the cause of this similar behaviour in the next section.

Before discussing consequences of CPT invariance in the next section, let us introduce a slightly more complicated top quark polarization asymmetries whose signs also measure the sign of $\xi$. 
We recall that the $t$ polarization perpendicular to the $W^+b\to th$ scattering plane $P_2$ can be expressed as a triple three-vector product\,(\ref{eq:P2_3product}), which is naive ${\rm T}$-odd (${\rm \tilde{T}}$-odd), since it changes the sign when we changes the signs of both the three momentum and spin. 
In the absence of final state re-scattering phase, ${\rm \tilde{T}}$-odd observables measure ${\rm T}$-violation, or CP-violation in quantum field theories (QFT). 
Therefore, we examine pentuple products
\begin{eqnarray}
\frac{(\vec{q}\times\vec{p}_j)\times(\vec{q}\times\vec{p}_h)\cdot\vec{P}}{|(\vec{q}\times\vec{p}_j)\times(\vec{q}\times\vec{p}_h)|},
\quad\quad 
\frac{(\vec{p}_b\times\vec{p}_j)\times(\vec{p}_b\times\vec{p}_h)\cdot\vec{P}}{|(\vec{p}_b\times\vec{p}_j)\times(\vec{p}_h\times\vec{p}_d)|},
\end{eqnarray}
which are clearly ${\rm \tilde{T}}$-odd polarization asymmetries, whose expectation values should vanish at $\xi=0$ in the tree level.
We note that the three-vector $(\vec{q}\times\vec{p_j})\times(\vec{q}\times\vec{p}_h)$ points toward the direction of $\vec{q}$, while its sign changes when the azimuthal angle between the $W^+$ emission plane and the $W^+b\to th$ scattering plane changes sign, between $-\pi<\phi<0$ and $0<\phi<\pi$. 
Likewise, $(\vec{p}_b\times\vec{p}_j)\times(\vec{p}_b\times\vec{p_h})$ points either along or opposite of $\vec{p}_b$ direction, depending on the same azimuthal angle between the two planes, because the $W^+$ momentum $\vec{q}$ and the $b$ momentum $\vec{p}_b$ are back to back in the frames which define the emission and the scattering planes, see Fig.\,\ref{fig:frame}. 
In the top quark rest frame, the two three-vectors, $\vec{q}$ and $\vec{p}_b$ span the scattering plane, which is chosen as the $x$-$z$ plane in our analysis. 
Therefore, if we define the azimuthal asymmetry of the top quark polarization vector as 
\begin{eqnarray}
P_k^A=P_k(\phi>0)-P_k(\phi<0),
\end{eqnarray}
where $P_k(\phi>0)$ and $P_k(\phi<0)$ denotes, respectively, the top quark polarization of events with $\phi>0$ and $\phi<0$, $P_1^A$ and $P_3^A$ are ${\rm \tilde{T}}$-odd. 
This is because the $x$- and $z$-axis vectors are linear combination of $\vec{q}$ and $\vec{p}_b$ in the $t$-rest frame.

We show in Fig.\,\ref{fig:pol5} all three polarization asymmetries, $P_k^A$ for $k=1,2,3$, for $pp\to thj$ events in the left two panels $(a), (b)$, and for $pp\to \bar{t}hj$ in the right panels.
The upper plots in Fig.\,\ref{fig:pol5} $(a), (c)$ are for ${\tt W}=400$ GeV, while the bottom plots $(b), (d)$ are for ${\tt W}=600$ GeV, both for ${\tt Q}>100$ GeV.
As expected, $P_1^A=P_3^A=0$ for the SM ($\xi=0$). We find that $P_3^A>0$ for $\xi=0.05\pi$ (dashed curves) and $0.1\pi$ (dash-dotted curves) in all the regions of $\cos\theta^\ast,{\tt W},$ and ${\tt Q}$ that we study, including the four cases shown in Fig.\,\ref{fig:pol5}.
This follows our observation that $P_3$ is large and opposite in sign between $t$ and $\bar{t}$, see Fig.\,\ref{fig:pol4}, and that azimuthal angle asymmetry is also opposite in sign, see Fig.\,\ref{fig:Aphi}. 
The magnitude of $P_1^A$ is small near $\cos\theta^\ast=-1$ where the cross section is large.


\section{${\rm T}$-odd v.s.\,${\rm CPV}$ asymmetry}\label{section:Todd}
\begin{figure}[h!]
\vspace{0.5cm}
\begin{centering}
\includegraphics[width=0.9\textwidth]{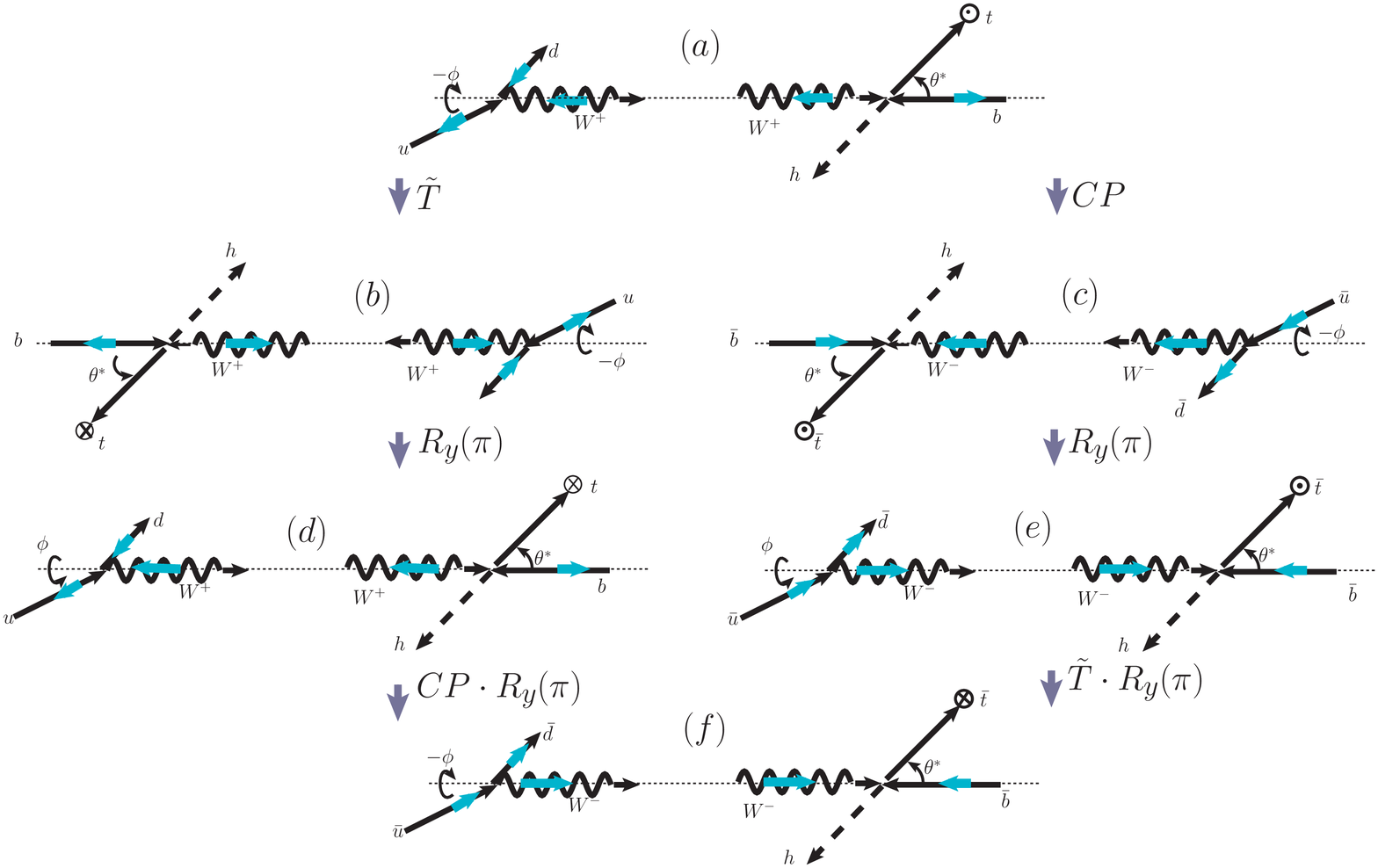}\label{fig:2}
\caption {Illustration of ${\rm \tilde{T}}$ and CP transformations of the process $ub\to dth$. 
$(a)$ shows the three momenta and polarization of $u\to d W^+$ and $W^+b\to th$ as parametrized in Fig.\,\ref{fig:frame}. 
The $W^+b\to th$ scattering is in the $z$-$x$ plane, $u$ and $d$ three momenta have negative $y$-components, and the top polarization is along the positive $y$-axis. 
$(b)$ is obtained from $(a)$ by ${\rm \tilde{T}}$ transformation, while $(c)$ is obtained from $(a)$ by CP transformation. $(d)$ and $(e)$ are obtained , respectively, from $(b)$ and $(c)$ by a $180$ degree rotation about the $y$-axis.
$(f)$ is obtained from $(d)$ by CP, or from $(e)$ by ${\rm \tilde{T}}$ transformation, together with the 180 degrees rotation about the $y$-axis.
}
\label{fig:lorentz}
\end{centering}
\end{figure}

As explained in the previous sections, the asymmetries $A_\phi$, $P_2$, $P_1^A$ and $P_3^A$, whose signs measure the sign of the CP violating phase $\xi$ are all so-called ${\rm T}$-odd asymmetries. 
We found in section \ref{sec:A_azimuthal} that the asymmetry $A_\phi$ has opposite sign between the $pp\to thj$ events and $pp\to\bar{t}hj$ events, and we found in section \ref{sec:pol} the polarization asymmetry $P_2$ has the opposite sign between the $thj$ and $\bar{t}hj$ events. In this section, we study consequences of the invariance under the discrete unitary transformations ${\rm \tilde{T}}$ and CP, and CP${\rm \tilde{T}}$.

 We adopt the symbol $\tilde {\rm T}$ for the unitary transformation under which all the three momenta $\vec{p}$ and the spin vectors $\vec{s}$ reverse their sign, in order to distinguish it from the time reversal transformation ${\rm T}$, which reverses the sign of the time direction, and hence is anti-unitary.
In the absence of the final state interaction phases of the amplitudes, ${\rm \tilde{T}}$-odd asymmetries are proportional to ${\rm T}$ violation, or equivalently CP violation in QFT.

Fig.\,\ref{fig:lorentz} illustrates the ${\rm \tilde{T}}$ and CP transformations of the subprocess $ub\to dth$, whose three momenta are the same as those in Fig.\,\ref{fig:frame}, or Eqs.\,(\ref{eq:pud}) and (\ref{eq:qpbptph1}). 
We add the helicities of external massless quarks ($u,\,d,\,b$) and also along the $W^+$ momentum direction, where the $\lambda=-1$ state is chosen for illustration. 
The top polarization, or its decay charged lepton momentum, is normal to the scattering plane along the positive $y$-axis. 
Under ${\rm \tilde{T}}$ transformation, all the three momenta and spin polarizations change sign, as shown in $(b)$, which can be viewed as $(d)$ by making the 180 degree rotation about the $y$-axis. Comparing $(a)$ and $(d)$, we find that the initial state remains the same while in the final state
\begin{eqnarray}
\phi\to-\phi \quad {\rm and} \quad P_2\to-P_2
\end{eqnarray}
under ${\rm \tilde{T}}$ transformation. Therefore the observation of ${\rm \tilde{T}}$-odd asymmetries such as 
\begin{eqnarray}
A_\phi\neq0\quad{\rm or}\quad
 P_2\neq0,
\end{eqnarray}
implies either ${\rm T}$-violation or the presence of an absorptive phase of the scattering amplitudes or both\,\cite{DeRujula:1978bz, Hagiwara:1982cq}.

Likewise, the configuration $(c)$ or $(e)$ after the $R_y(\pi)$ rotation, is obtained by CP transformation from the configuration $(a)$. 
All the particles are transformed to anti-particles and their helicities and three momenta are reversed. 
If we define the asymmetries $\overline{A}_\phi$ and $\overline{P}_2$ for the process $\bar{p}\bar{p}\to\bar{t}hj$, then CP-invariance between $(a)$ and $(e)$ implies
\begin{eqnarray}\label{eq:Aphibar_P2bar}
\overline{A}_\phi = -A_\phi \quad {\rm and}\quad
  \overline{P}_2 = P_2  \,.
\end{eqnarray}
Violation of the above identities hence gives CP-violation. 

Finally, the configuration $(f)$ in Fig.\,\ref{fig:lorentz} is obtained from $(d)$ by applying CP, or from $(e)$ by applying ${\rm \tilde{T}}$, together with the rotation $R_y(\pi)$.
 In short, $(f)$ is obtained from our original configuration $(a)$ by CP${\rm \tilde{T}}$ transformation\,\cite{Hagiwara:1986vm}. By comparing $(a)$ and $(f)$, CP${\rm \tilde{T}}$ invariance, or the absence of the absorptive phase in QFT amplitudes should give 
\begin{eqnarray}\label{eq:Aphibar=Aphi}
  \overline{A}_\phi = A_\phi \quad {\rm and}\quad
  \overline{P}_2 = -P_2                
\end{eqnarray}

\begin{figure}[t]
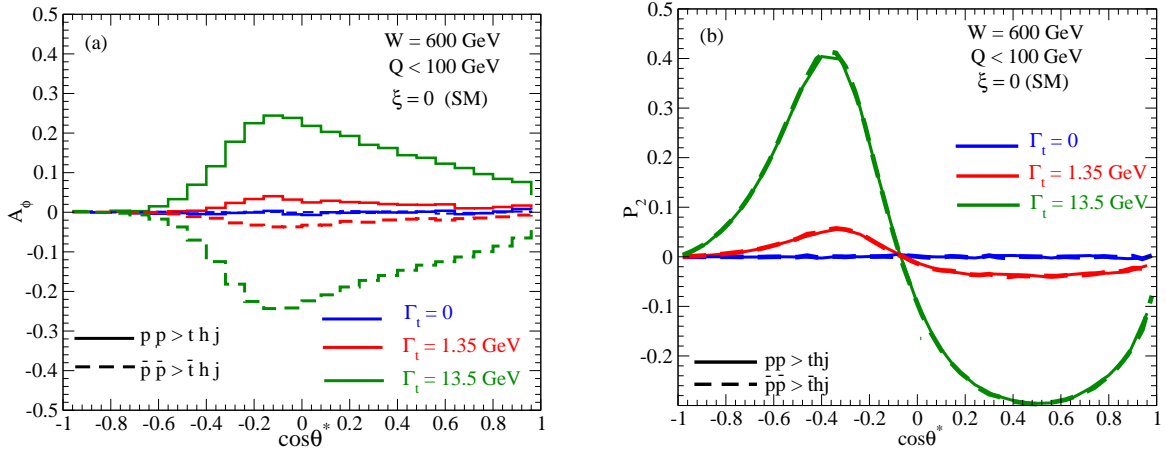

\vspace{0.5cm}
\begin{centering}
\includegraphics[width=0.4\textwidth]{Aphi_pppbpb.eps}\label{}
\hspace{0.8cm}
\includegraphics[width=0.4\textwidth]{P2_Gt.eps}\label{}
\caption {${\rm \tilde{T}}$-odd asymmetries $A_\phi$ (a) and $P_2$ (b) due to the top quark width $\Gamma_t$ in CP invariant theory ($\xi=0$) in $pp\to thj$ (solid lines) and $\bar{p}\bar{p}\to\bar{t}hj$ (dashed lines) at {\tt W} = 600 GeV for events with ${\tt Q}<100$ GeV. Three values of $\Gamma_t$ are shown: $\Gamma_t=0$ (blue), $\Gamma_t=1.35$ GeV (red), and ten times the SM value $\Gamma_t=13.5$ GeV (green).}
\label{fig:AphiP2_Gt}
\end{centering}
\end{figure}
As an illustration of how absorptive phases of the amplitudes in ${\rm T}$ or CP invariant theory contribute to ${\rm \tilde{T}}$-odd asymmetries, we examine the impacts of the top-quark width in the $s$-channel propagator $D_t(P_{th})$ in Eq.\,(\ref{eq:Tnu}), or in the $B$ factor of Eq.\,(\ref{eq:B}). 
The width of Breit-Wigner propagator gives absorptive parts to our amplitudes, and since the top quark width appears only in the amplitudes with $htt$ coupling, it can give rise to ${\rm \tilde{T}}$-odd asymmetries, $A_\phi$ and $P_2$. 
We show in Fig.\,\ref{fig:AphiP2_Gt} the asymmetries $A_\phi$ $(a)$ and $P_2$ $(b)$ in the CP-invariant SM ($\xi=0$) for $\Gamma_t=0$ (blue), $\Gamma_t=1.35$ GeV (red), the SM value, and for 10 times the SM width $\Gamma_t=13.5$ GeV (green).
We find that the 
asymmetries are both zero when $\Gamma_t = 0$ as expected. 
Furthermore, we confirm the relations\,(\ref{eq:Aphibar_P2bar}) between the asymmetries of $pp\to thj$ events, $A_\phi$ and $P_2$, and those of $\bar{p}\bar{p}\to\bar{t}hj$ events, $\overline{A}_\phi$ and $\overline{P}_2$, respectively.
This is a consequence of CP invariance, as can be viewed 
from the illustration by comparing the configurations $(a)$ 
and $(e)$.
If CP is conserved, the amplitudes for the configuration 
$(e)$ should have the same magnitude with those of the 
original configuration $(a)$.
The azimuthal angle between the $W$ emission plane and 
the scattering plane is reversed, whereas the $\bar{t}$ 
spin polarization should be the same as the $t$ spin 
polarization.

It is worth noting here that instead of top and anti-top 
spin polarization vector, $P_2$ and 
$\overline{P}_2$, if we use the decay charged lepton 
momentum normal to the scattering plane in the $t$ or 
$\bar{t}$ rest frame, we find
\begin{eqnarray}
\label{eq:pl+-}
  \langle p^{l^-}_y\rangle = -\langle p^{l^+}_y\rangle           
\end{eqnarray}
as a consequence of $\overline{P}_2=P_2$\,(\ref{eq:Aphibar_P2bar}) in CP invariant theory.
Here, we assume that the $t$ and $\bar{t}$ decay 
angular distributions follow the SM, where the charged 
leptons are emitted preferably along the $t$ spin 
polarization direction, whereas they are emitted in 
the opposite of the $\bar{t}$ spin polarization 
direction.
This is simply because only the right-handed $l^+$ 
and the left-handed $l^-$ are emitted from $t$ 
and $\bar{t}$ decays, respectively, in the SM. 
The above spin-momentum correlation is CP invariant, 
and hence the identity (\ref{eq:pl+-}) is also a consequence 
of CP invariance.

\begin{figure}[b]
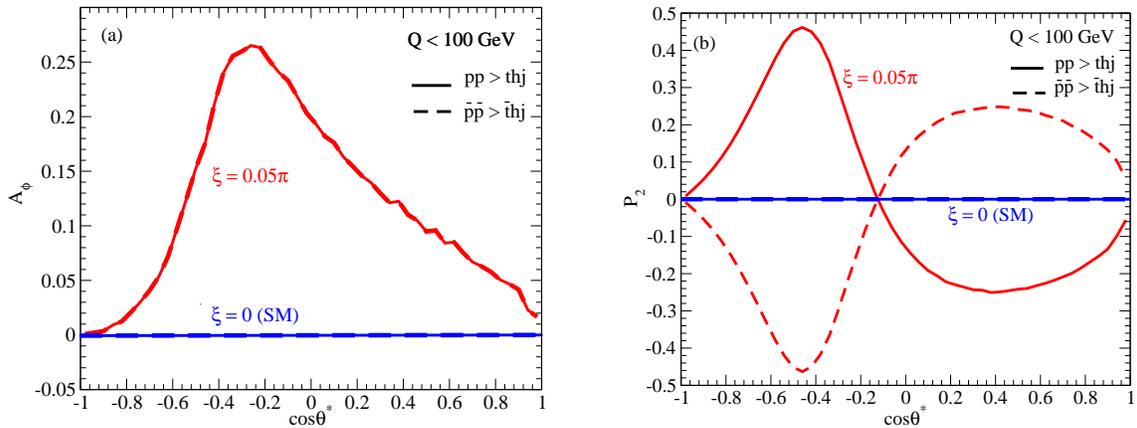

\begin{centering}
\vspace{0.5cm}
\includegraphics[width=0.4\textwidth]{Aphi_costt_PDFcheck_Qlt100.eps}\label{}
\hspace{0.8cm}
\includegraphics[width=0.38\textwidth]{P2_PDFcheck_Qlt100.eps}\label{}
\caption {The azimuthal angle asymmetry $A_\phi$ $(a)$ and the top (anti-top) polarization asymmetry $P_2$ $(b)$ in $pp\to thj$ events (solid curves) and $\bar{p}\bar{p}\to\bar{t}hj$ events (dashed curve) for $\xi=0$ (SM, blue curve), and $\xi=0.05\pi$ (red curve).  }
\label{fig:AphiP2_ppbar}
\end{centering}
\end{figure}

In Fig.\,\ref{fig:AphiP2_ppbar}, we show comparisons of the asymmetries 
between $pp \to thj$ and $\bar{p}\bar{p}\to \bar{t}hj$ 
events for CP violating theory ($\xi \neq 0$) in 
the approximation of no absorptive parts in the 
amplitudes, i.e., we set $\Gamma_t=0$. 
We confirm the relations\,(\ref{eq:Aphibar=Aphi}) 
for the same value of $\xi$, as a consequence of CP${\rm \tilde{T}}$ invariance. 
The relations between the asymmetries in $pp\to thj$ 
and $\bar{p}\bar{p}\to \bar{t}hj$ are opposite between 
Fig.\,\ref{fig:AphiP2_Gt} and Fig.\,\ref{fig:AphiP2_ppbar}, as expected from Eqs.\,(\ref{eq:Aphibar_P2bar}) and\,(\ref{eq:Aphibar=Aphi}).

All the above relations between $pp$ and $\bar{p}\bar{p}$ may seem to be just formal rules 
since we will not have a $\bar{p}\bar{p}$ collider with the LHC energy and luminosity. 
However, we find the above rules useful in testing our amplitudes, 
especially in fixing the relative sign between the two 
helicity amplitudes which determines the top and anti-top 
spin polarization directions away from their helicity axis.
Furthermore, we find that it is possible to disentangle 
${\rm \tilde{T}}$-odd effects coming from the SM re-scattering 
effects (that give rise to the absorptive amplitudes) 
from CP violating new physics effects in $pp$ collisions 
at the LHC by measuring the polarization asymmetry $P_2$ 
of $t$ and $\bar{t}$ precisely.

Let us examine Fig.\,\ref{fig:P2_W} again, where we show $P_2$ for 
$thj$ and $\bar{t}hj$ events at the LHC as a function of ${\tt W}$, 
the $th$ or $\bar{t}h$ invariant mass. 
The polarization asymmetry $P_2$ have opposite sign 
between $t$ and $\bar{t}$. 
More quantitatively, we note that the magnitudes of 
the asymmetry is almost the same for small ${\tt Q}$ events 
(${\tt Q}<$~100~GeV) at large ${\tt W}$ (${\tt W} \gtrsim$ 600~GeV). 
This is a consequence of CP${\rm \tilde{T}}$ invariance 
of our tree-level amplitudes with $\Gamma_t=0$, 
because at small ${\tt Q}$ and large ${\tt W}$, the events are 
dominated by the contributions of the longitudinally 
polarized $W$ bosons; see Fig.\,\ref{fig:QW}\,(a) and\,(b). 
Therefore, in this region of the phase space, 
we can regard the single top or anti-top plus Higgs production 
processes as 
\begin{subequations}
\begin{align}
 & W^+(\lambda=0) + b \to t + h                 
\\
 & W^-(\lambda=0) + \bar{b} \to \bar{t} + h     
\end{align}
\end{subequations}
which are CP conjugates of each other. 
Their amplitudes are given in Eqs.\,(\ref{eq:T0}) and\,(\ref{eq:CPtransform}), and 
we can obtain the polatization asymmetries directly 
from these amplitudes, which are independent of 
parton distribution functions in $pp$ collisions.

Because the absorptive amplitudes contribute to the 
polarization asymmetry $P_2$ with the same sign as 
shown in Fig.\,\ref{fig:AphiP2_Gt}, we can further tell that the 
difference, 
\begin{eqnarray}
  P_2~(thj ~{\rm events})
-P_2~( \bar{t}hj~ {\rm events})          
\end{eqnarray}
measures CP violation, whereas the sum 
\begin{eqnarray}
  P_2~( thj~ {\rm events})
+P_2~( \bar{t}hj~ {\rm events})         
\end{eqnarray}
measures the CP${\rm \tilde{T}}$-odd effects from the 
absorptive amplitudes in the region of small ${\tt Q}$ 
and large ${\tt W}$. 
We find in the SM the leading contributions for 
the absorptive amplitudes appear at one-loop level 
in QCD and in the electroweak 
theory\,\cite{Barger:2020}. 
The top quark width that we adopted in this section for 
illustration is a part of the electroweak corrections.

The sign of the polarization asymmetry $P_2$ remains 
the same and the magnitudes are larger at smaller ${\tt W}$ 
and large ${\tt Q}$. 
This can be understood qualitatively also from Fig.\,\ref{fig:QW}, 
where the sub-dominant contributions are 
\begin{subequations}
\begin{align}
&  W^+(\lambda=-1) + b \to t + h      \\           
 & W^-(\lambda=+1) + \bar{b} \to \bar{t} + h    
  \end{align}
\end{subequations}
at small ${\tt W}$ (${\tt W} \lesssim 500$~GeV) especially at 
large ${\tt Q}$ (${\tt Q} > 100$~GeV). 
The above subprocesses are again CP-conjugate to 
each other, and hence follow the rule\,(\ref{eq:Aphibar=Aphi}) from 
CP${\rm \tilde{T}}$ invariance.

\section{Summary and discussions}
\label{sec:summary}

We studied associated production of single top (or anti-top) 
and the Higgs boson via $t$-channel $W$ exchange at the LHC. 
We obtained analytically the helicity amplitudes for all 
the tree-level subprocesses with massless $b$ (or $\bar{b}$) 
quark PDF in the proton, and studied consequences of possible 
CP violation in the Higgs Yukawa coupling to the top quark. 
By choosing the momentum direction of the $W^\pm$ exchanged 
in the $t$-channel, the helicity amplitudes are factorized 
into the $W^\pm$ emission amplitudes from light quarks or
anti-quarks, and the $W^+b\to th$ or $W^-\bar{b}\to\bar{t}h$ 
production amplitudes. 
We find that the amplitudes for the right-handed top quark 
and those of the left-handed anti-top quark are sensitive 
to the sign of the CP violating phase $\xi$ in the 
effective Yukawa interaction Lagrangian of Eq.\,(\ref{eq:coupling}).
This is because the right-handed top quark is produced by the $t_R^\dagger t_L$  operator with the $e^{-i\xi}$ phase without chirality suppression, whereas the contribution of the $t_L^\dagger t_R$ operator with the $e^{i\xi}$ phase is doubly suppressed. For the anti-top production, the role of the two operators are reversed. 
On the other hand, the other amplitudes for the left-handed 
top and the right-handed anti-top productions are almost 
proportional to $e^{i\xi}+e^{-i\xi}=2\cos\xi$ because 
both terms in the Lagrangian contribute with one chirality 
suppression, either in the top quark propagator or from the 
helicity-chirality mismatch in the wave function, $\delta^\prime$ and $\delta$ in Eq.\,(\ref{eq:deltadeltaprime}), respectively.

We studied mainly the azimuthal angle asymmetry $A_\phi$ 
between the $W^\pm$ emission plane and the $W^+ b \to t h$ or 
$W^- \bar{b} \to \bar{t} h$ production plane, and the 
$t$ or $\bar{t}$ spin polarization normal to the 
scattering plane, $P_2$, as observables which are sensitive 
to the sign of the CP phase $\xi$. 
The asymmetry $A_\phi$ arises from the interference between 
the amplitudes with longitudinal and transversely polarized 
$W^\pm$ contributions, and hence is significant when the 
exchanged momentum transfer ${\tt Q}$ is relatively large and 
the $th$ or $\bar{t}h$ invariant mass ${\tt W}$ is not too large, 
where both of the interfering amplitudes are significant. 
The magnitude of the asymmetry can be enhanced by selecting 
the chirality favored top or anti-top quark helicity, 
e.g.\,by selecting those events with charged lepton momentum 
along the top or anti-top momentum direction in the $th$ 
or $\bar{t}h$ rest frame; see Fig.\,\ref{fig:Aphi}.

On the other hand, the polarization asymmetry $P_2$ is 
obtained as the interference between the two helicity 
amplitudes of $t$ or $\bar{t}$. 
We find that the amplitudes are dominated by the collision 
of longitudinally polarized $W^\pm$ and $b$ or $\bar{b}$ 
when the momentum transfer ${\tt Q}$ is small and the invariant 
mass ${\tt W}$ of the $th$ or $\bar{t}h$ system is large. 
Therefore in such kinematical configuration, 
the asymmetry $P_2$ of the top and the anti-top 
can be regarded as the direct test of CP violation 
between the CP-conjugate processes, 
$W^+(\lambda=0) +b \to t+h$ and 
$W^-(\lambda=0)+ \bar{b} \to \bar{t}+h$. 
Because of the dominance of the longitudinally polarized 
$W^\pm$ exchange amplitudes, all the differences in 
the quark and anti-quark PDF's of the colliding protons 
drop out in the polarization asymmetry.

All the analytic and numerical results presented in this 
report are done strictly in the tree-level, in order to 
clarify the symmetry properties of observable asymmetries 
that are sensitive to the sign of the CP violating phase 
$\xi_{htt}$. 
In order to show their observability at the HL-LHC with 
its 3 ab$^{-1}$ of integrated luminosity, we should perform 
the following studies.

Most importantly, we should identify the top and the Higgs 
decay modes which can be used to measure the asymmetries, 
since we may have 
different radiative corrections and background contributions 
for each set of the decay modes. 
We expect that semi-leptonic decays of $t$ and 
$\bar{t}$ when the Higgs decays into modes without missing 
energy are favorable because the lepton charge identify $t$ vs.\,$\bar{t}$, and the charged lepton decay 
anglular distribution measures the $t$ and $\bar{t}$ 
polarization with maximum sensitivity. 
Hadronically decaying $t$ and $\bar{t}$ events can have 
sensitivity to the asymmetries, because their decay 
density matrix polarimeter introduced in Ref.\,\cite{Hagiwara:2017ban} 
retains strong sensitivity to the $t$ and $\bar{t}$ 
polarizations, and also because the CP asymmetry of the 
polarizations, $P_2(\bar{t}hj) \approx -P_2(thj)$ in Fig.\,\ref{fig:P2_W}
tells that the observable asymmetries in the decay 
distributions are the same between $t$ and $\bar{t}$ 
events even if we cannot distinguish between them. 
Although the direct test of CP violation cannot be made 
in the hadronic decay modes, the sensitivity to the sign 
and the magnitude of the CP violating phase $\xi$ can 
be improved by assuming the SM radiative contribution 
to the asymmetries\,\cite{Barger:2020}.

We believe that the associated production of the Higgs 
boson and single $t$ or $\bar{t}$ via $t$-channel $W^\pm$ 
exchange at the LHC can be an ideal testing ground of 
the top quark Yukawa coupling, because the amplitudes 
with the $htt$ Yukawa coupling and those of the $hWW$ 
coupling interfere strongly. 
We studied the sensitivity of the process to possible 
CP violation in the Yukawa coupling. 
We anticipate that our studies based on the analytic form 
of the helicity amplitudes will be useful in the test of various 
scenarios of physics beyond the SM.

\appendix
\section{Helicity amplitudes and the top spin orientation}
\label{sec:AppA}
The helicity amplitudes $M_+$ and $M_-$ in the single top plus Higgs production processes via $t$-channel $W$ exchange are pure complex numbers when all the other quark masses are set to be zero, because all their helicities are fixed by the SM $V-A$ interactions and because the Higgs boson has no spin. 
The produced top quark is hence a pure quantum state
\begin{eqnarray}
\label{eq:App_top}
|t\rangle=\frac{M_+}{\sqrt{|M_+|^2+|M_-|^2}}
\left(
{\begin{array}{*{20}{c}}
1\\
0
\end{array}}
\right)
+
\frac{M_-}{\sqrt{|M_+|^2+|M_-|^2}}
\left(
{\begin{array}{*{20}{c}}
0\\
1
\end{array}}
\right)
\end{eqnarray}
in the top rest frame. Its spin is polarized in the positive $z$-direction, $J_z=\frac{1}{2}$,
if $M_+\neq0$ and $M_-=0$, where the $z$-axis is along the top quark momentum direction where its helicity is defined. 
If $M_+=0$ but $M_-\neq0$, the top quark has $J_z=-\frac{1}{2}$.
In general, the spin polarization of the top quark can have an arbitrary spatial orientation
\begin{eqnarray}
\label{eq:App_J}
\vec{J}=\frac{1}{2}\vec{P}=\frac{1}{2}(\sin\theta\cos\phi,\sin\theta\sin\phi,\cos\theta),
\end{eqnarray}
where $\theta$ and $\phi$ are polar and azimuthal angles about the $z$-axis. 
The corresponding top state can be obtained from the $\vec{P}=(0,0,1)$ state by two rotations,
\begin{eqnarray}
\label{eq:App_decomposition}
|t,\vec{P}(\theta,\phi)\rangle
&=&
R_z(\phi)R_y(\theta)
\left(
{\begin{array}{*{20}{c}}
1\\
0
\end{array}}
\right)
\nonumber\\
&=&
\left(
{\begin{array}{*{20}{cc}}
e^{-i\frac{\phi}{2}}&0\\
0&e^{i\frac{\phi}{2}}
\end{array}}
\right)
\left(
{\begin{array}{*{20}{cc}}
\cos\frac{\theta}{2}&-\sin\frac{\theta}{2}\\
\sin\frac{\theta}{2}&\cos\frac{\theta}{2}
\end{array}}
\right)
\left(
{\begin{array}{*{20}{c}}
1\\
0
\end{array}}
\right)
\nonumber\\
&=&
e^{-i\frac{\phi}{2}}\cos\frac{\theta}{2}
\left(
{\begin{array}{*{20}{c}}
1\\
0
\end{array}}
\right)
+
e^{i\frac{\phi}{2}}\sin\frac{\theta}{2}
\left(
{\begin{array}{*{20}{c}}
0\\
1
\end{array}}
\right)
\end{eqnarray}
By comparing\,(\ref{eq:App_top}) and\,(\ref{eq:App_decomposition}), we find 
\begin{eqnarray}
\frac{M_-}{M_+}={e^{i\phi}}{\tan\frac{\theta}{2}}
\end{eqnarray}
or 
\begin{eqnarray}
\theta=2\tan^{-1}\left|\frac{M-}{M_+}\right|,
\quad 
\phi=\arg\left(\frac{M_-}{M_+}\right).
\end{eqnarray}

For the mixed states, it is useful to introduce the density matrix 
\begin{eqnarray}
\rho_{\sigma\sigma^\prime}^{}
=
\frac{1}{\sum(|M_+|^2+|M_-|^2)}
\left(
{\begin{array}{*{20}{cc}}
\sum|M_+|^2&\sum M_+M_-^\ast\\
\sum M_+^\ast M_-&\sum|M_-|^2
\end{array}}
\right)
\end{eqnarray}
where the summation is over all the processes and kinematical configurations that contribute to the top quark which we observe.
Because the matrix is Hermitian and has trace 1, we can parametrize it as 
\begin{eqnarray}
\label{eq:App_rho}
\rho=\frac{1}{2}\left(1+\vec{P}\cdot\vec{\sigma}\right)
\end{eqnarray}
by using the $\vec{\sigma}$ matrices. 
We find
\begin{eqnarray}
P_1=\frac{2Re\sum(M_+M_-^\ast)}{\sum(|M_+|^2+|M_-|^2)},
\quad\quad
P_2=\frac{-2Im\sum(M_+M_-^\ast)}{\sum(|M_+|^2+|M_-|^2)},
\quad\quad
P_3=\frac{\sum(|M_+|^2-|M_-|^2)}{\sum(|M_+|^2+|M_-|^2)},
\end{eqnarray}
which for the pure state\,(\ref{eq:App_top}) gives\,(\ref{eq:App_J}).

In general, we can parametrize the density matrix\,(\ref{eq:App_rho}) as 
\begin{eqnarray}
\rho=\frac{1-|\vec{P}|}{2}
+
\frac{|\vec{P}|}{2}
\left(
1+\frac{\vec{P\cdot\vec{\sigma}}}{|\vec{P}|}
\right),
\end{eqnarray}
which is a sum of unpolarized top quark with the probability $1-|\vec{P}|$, and the fully polarized top quark with its spin polarization orientation along
\begin{eqnarray}
\label{eq:App_Ptheta}
\vec{P}=|\vec{P}|(\sin\theta\cos\phi,\cos\theta\sin\phi,\cos\theta)
\end{eqnarray}
with the probability $|\vec{P}|$. 
We find it convenient to show the general polarization vector $\vec{P}$
(\ref{eq:App_Ptheta}) by using  an arrow of length $|\vec{P}|$ in the polar coordinate defined as 
\begin{eqnarray}
-\pi<\theta\leq\pi,\quad\quad\frac{-\pi}{2}<\phi\leq\frac{\pi}{2}.
\end{eqnarray}
When the imaginary part of $M_-/M_+$ is small, we tend to have small $|\phi|$, and with the above definition we can show $\phi>0$ and $\phi<0$ as pointing up and down in the $z$-$x$ plane\,\cite{Barger:2018tqn} .

%
%
\section{Polarized $t$ and $\bar{t}$ decay distributions}
\label{sec:AppB}
The general mixed state of $t$ and $\bar{t}$ in a given kinematical configuration is described by the polarization density matrix (\ref{eq:App_rho}) with the polarization vector, $\vec{P}=(P_1,P_2,P_3)$ with $|\vec{P}|<$ 1. In this appendix we give $t$ and $\bar{t}$ decay angular distributions for both semi-leptonic and hadronic decay modes.

The decay density matrix for semi-leptonic decay modes is very simple because it depends only on the charged lepton polar and azimuthal angles\,\cite{Atwood:1992vj, Atwood:2000tu,Hagiwara:2017ban}
\begin{subequations}
\begin{align}
d\rho(t\to b\ell^+\nu)
&=
B(t\to b\ell\nu)
\left(
{\begin{array}{*{20}{cc}}
1+\cos\bar{\theta}^\ast&\sin\bar{\theta}^\ast e^{i\bar{\phi}^\ast}\\
\sin\bar{\theta}^\ast e^{-i\bar{\phi}^\ast}&1-\cos\bar{\theta}^\ast
\end{array}}
\right)
\frac{d\cos\bar{\theta}^\ast d\bar{\phi}^\ast}{4\pi}
\label{eq:AppB_drhot}
\\
d\rho(\bar{t}\to \bar{b}\ell^-\bar{\nu})
&=
B(t\to b\ell\nu)
\left(
{\begin{array}{*{20}{cc}}
1-\cos{\theta}^\ast&\sin{\theta}^\ast e^{-i{\phi}^\ast}\\
\sin{\theta}^\ast e^{i\bar{\phi}^\ast}&1+\cos{\theta}^\ast
\end{array}}
\right)
\frac{d\cos{\theta}^\ast d{\phi}^\ast}{4\pi}
\label{eq:AppB_drhotbar}
\end{align}
\end{subequations}
where the $\ell^\pm$ four momenta in the $t$ and $\bar{t}$ rest frame, respectively, are parametrized as 
\begin{subequations}
\begin{align}
p_{\ell^+}^\mu&=\frac{m_t}{2}\bar{x}(1,\sin\bar{\theta}^\ast\cos\bar{\phi}^\ast,\sin\bar{\theta}^\ast\sin\bar{\phi}^\ast,\cos\bar{\theta}^\ast),\\
p_{\ell^-}^\mu&=\frac{m_t}{2}{x}(1,\sin{\theta}^\ast\cos{\phi}^\ast,\sin{\theta}^\ast\sin{\phi}^\ast,\cos{\theta}^\ast).
\end{align}
\end{subequations}
The $z$-axis is along the $t$ or $\bar{t}$ helicity axis and the $y$-axis is along the normal to the scattering plane, $\vec{q}\times\vec{p}_t$ or $\vec{q}\times\vec{p}_{\bar t}$, respectively, where the helicity amplitudes are calculated
\footnote{
Note that the $\bar{t}\to\bar{b}\ell^-\bar{\nu}$ decay density matrix distributions given in Eq.\,(A30) of Ref.\,\cite{Hagiwara:2017ban} differs from\,(\ref{eq:AppB_drhotbar}), because the reference frame in Ref.\,\cite{Hagiwara:2017ban} has been chosen to have common $z$- and $x$-axis for both $t\to b\ell^+\nu$ and $\bar{t}\to\bar{b}\ell^-\bar{\nu}$ decays in the $t\bar{t}$ rest frame in order to study $t$ and $\bar{t}$ decay angular correlations effectively in the process $e^+e^-\to ht\bar{t}$.
}.
The decay angular distributions are then 
\begin{subequations}
\begin{align}
d\Gamma(t(\vec{P})\to b\bar{\ell}\nu)
&=
\sum_{\sigma}\sum_{\sigma^\prime}\rho_{\sigma\sigma^\prime}^t(\vec{P}) d\rho(t\to b\bar{\ell}\nu)_{\sigma\sigma^\prime}
\nonumber
\\
&=
B(t\to b\bar{\ell}\nu)
\left\{
1+P_1\sin\bar{\theta}^\ast\cos\bar{\phi}^\ast+P_2\sin\bar{\theta}^\ast\sin\bar{\phi}^\ast+P_3\cos\bar{\theta}^\ast
\right\}
\frac{d\cos\bar{\theta}^\ast d\bar{\phi}^\ast}{4\pi},
\\
d\Gamma(\bar{t}(\vec{P})\to \bar{b}\ell\bar{\nu})
&=
\sum_{\sigma}\sum_{\sigma^\prime}\rho_{\sigma\sigma^\prime}^{\bar{t}}(\vec{P}) d\rho(\bar{t}\to \bar{b}\ell\bar{\nu})_{\sigma\sigma^\prime}
\nonumber
\\
&=
B(t\to b\bar{\ell}\nu)
\left\{
1-P_1\sin{\theta}^\ast\cos{\phi}^\ast-P_2\sin{\theta}^\ast\sin{\phi}^\ast-P_3\cos\bar{\theta}^\ast
\right\}
\frac{d\cos{\theta}^\ast d\phi^\ast}{4\pi}.
\end{align}
\end{subequations}

The $t$ and $\bar{t}$ decay density matrix distributions for hadronic decay modes are slightly more complicated because it is difficult to identify the down-type quark jet from the up-type quark jet in the $W^+\to\bar{d}u(\bar{s}c)$ and $W^-\to d\bar{u}(s\bar{c})$ dijet system.
By assuming that the $b$ and $\bar{b}$ jet can be identified uniquely,
\begin{subequations}
\begin{align}
p_b^\mu&=\frac{m_t}{2}x_b(1,\sin\theta^\ast_b\cos\phi^\ast_b,\sin\theta^\ast_b\sin\phi^\ast_b,\cos\theta^\ast_b),
\\
p_{\bar{b}}^\mu&
=
\frac{m_t}{2}x_{\bar{b}}(1,\sin\theta^\ast_{\bar{b}}\cos\phi^\ast_{\bar{b}},\sin\theta^\ast_{\bar{b}}\sin\phi^\ast_{\bar{b}},\cos\theta^\ast_{\bar{b}}),
\end{align}
\end{subequations}
in the $t$ and $\bar{t}$ rest frame, respectively, with $x_b=x_{\bar{b}}=1-m_W^2/m_t^2$ in the narrow $W$ width approximation, the $\bar{d}$ and $u$ four momenta are parametrized in the $W^+\to d\bar{u}$ rest frame as
\begin{subequations}\label{eq:AppB_pdbaru}
\begin{align}
p_{\bar{d}}^\mu
&=
\frac{m_W^{}}{2}
(1,
\sin\bar{\theta}^{\ast\ast}\cos\bar{\phi}^{\ast\ast}, 
\sin\bar{\theta}^{\ast\ast}\sin\bar{\phi}^{\ast\ast}, 
\cos\bar{\theta}^{\ast\ast}
)
\label{eq:AppB_pdbar}
\\
p_{u}^\mu
&=
\frac{m_W^{}}{2}
(1,
-\sin\bar{\theta}^{\ast\ast}\cos\bar{\phi}^{\ast\ast}, 
-\sin\bar{\theta}^{\ast\ast}\sin\bar{\phi}^{\ast\ast}, 
-\cos\bar{\theta}^{\ast\ast}
).
\label{eq:AppB_pu}
\end{align}
\end{subequations}
Likewise, the $d$ and $\bar{u}$ four momenta are parametrized in the $W^-\to d\bar{u}$ rest frame as 
\begin{subequations}\label{eq:AppB_pdubar}
\begin{align}
p_d^\mu&=
\frac{m_W}{2}
(1,
\sin\theta^{\ast\ast}\cos\phi^{\ast\ast},
\sin\theta^{\ast\ast}\sin\phi^{\ast\ast},
\cos\theta^{\ast\ast}
),
\label{eq:AppB_pd}
\\
p_{\bar{u}}^\mu&=
\frac{m_W}{2}
(1,
-\sin\theta^{\ast\ast}\cos\phi^{\ast\ast},
-\sin\theta^{\ast\ast}\sin\phi^{\ast\ast},
-\cos\theta^{\ast\ast}
).
\end{align}
\end{subequations}

In the $t$ or $\bar{t}$ rest frame, the $\bar{d}$ or $d$ four momenta are obtained from\,(\ref{eq:AppB_pdbar}) or\,(\ref{eq:AppB_pd}), respectively, by a boost and can be expressed as 
\begin{subequations}
\begin{align}
p_{\bar{d}}^\mu
&=
\frac{m_t}{2}(\bar{x},\bar{x}_1,\bar{x}_2,\bar{x}_3),
\\
p_{{d}}^\mu
&=
\frac{m_t}{2}({x},{x}_1,{x}_2,{x}_3).
\label{eq:AppB_pdbarBoost}
\end{align}
\end{subequations}
In Ref.\,\cite{Hagiwara:2017ban}, it has been assumed that with the probability $P_{\bar{d}u}\geq0.5$, the $\bar{d}$-quark is correctly identified and with the probability $1-P_{\bar{d}u}\leq0.5$, the $u$-quark is mistaken as the $\bar{d}$ quark.
The $t\to b\bar{d}u$ decay density matrix distribution is then expressed as 
\begin{eqnarray}
d\rho(t\to b\bar{d}u)
=
\frac{6B(t\to b\bar{d}u)}{\left(1-\frac{m_W^2}{m_t^2}\right)\left(1+2\frac{m_W^2}{m_t^2}\right)}
\left[
\frac{1+P_{\bar{d}u}}{2}\hat{\rho}_{\bar{d}}+\frac{1-P_{\bar{d}u}}{2}\hat{\rho}^\prime_{\bar{d}}
\right]
\frac{d\cos\theta^\ast_b d\phi_b^\ast}{4\pi}
\frac{d\cos\bar{\theta}^{\ast\ast}d\bar{\phi}^{\ast\ast}}{4\pi}
\end{eqnarray}
where
\begin{eqnarray}\label{eq:AppB_rhohat}
\hat{\rho}_{\bar{d}}^{}
=
(1-\bar{x})
\left(
{\begin{array}{*{20}{cc}}
\bar{x}+\bar{x}_3&\bar{x}_1+i\bar{x}_2\\
\bar{x}_1-i\bar{x}_2&\bar{x}-\bar{x}_3
\end{array}}
\right),
\end{eqnarray}
and $\hat{\rho}^\prime_{\bar{d}}$ is obtained from\,(\ref{eq:AppB_rhohat}) by replacing the $\bar{d}$ and $u$  four momentum\,(\ref{eq:AppB_pdbaru}) in the $W^+$ rest frame. This simple density matrix distribution reduces to the charged lepton distribution\,(\ref{eq:AppB_drhot}) in the $P_{\bar{d}u}=1$ limit.
The decay density distribution for $\bar{t}\to\bar{b}\ell\bar{\nu}$ is obtained similarly as 
\begin{eqnarray}
d\rho(\bar{t}\to\bar{b}\ell\bar{\nu})
=
\frac{6B(\bar{t}\to\bar{b}d\bar{u})}
{\left(1-\frac{m_W^2}{m_t^2}\right)\left(1+2\frac{m_W^2}{m_t^2}\right)}
\left[
\frac{1+P_{d\bar{u}}}{2}\hat{\rho}_d
+\frac{1-P_{d\bar{u}}}{2}\hat{\rho}_d^\prime
\right]
\frac{d\cos\theta_{\bar{b}}^\ast d\phi_{\bar{b}}^\ast}{4\pi}
\frac{d\cos\theta^{\ast\ast}d\phi^{\ast\ast}}{4\pi}
\end{eqnarray}
where the density matrix
\begin{eqnarray}
\hat{\rho}_d
=
(1-x)
\left(
{\begin{array}{*{20}{cc}}
{x}-{x}_3&{x}_1-i{x}_2\\
{x}_1+i{x}_2&{x}+{x}_3
\end{array}}
\right)
\end{eqnarray}
is obtained from the $d$-quark momentum\,(\ref{eq:AppB_pdbarBoost}) in the $t$-rest frame, and $\hat{\rho}_d^\prime$ is obtained by exchanging the $d$ and $\bar{u}$ four momenta\,(\ref{eq:AppB_pdubar}) in the same event.

The decay angular distribution of arbitrary polarized $t$ and $\bar{t}$ are then obtained simply by taking the $'$trace$'$
\begin{subequations}\label{eq:AppB_dGamma}
\begin{align}
d\Gamma(t(\vec{P})\to b\bar{d}u)
&=
\sum_{\sigma}\sum_{\sigma^\prime}
\rho_{\sigma\sigma^\prime}^t(\vec{P})
d\rho(t\to b\bar{d}u)_{\sigma\sigma^\prime},
\label{eq:AppB_dGammat}
\\
d\Gamma(\bar{t}(\vec{P})\to \bar{b}{d}\bar{u})
&=
\sum_{\sigma}\sum_{\sigma^\prime}
\rho_{\sigma\sigma^\prime}^{\bar{t}}(\vec{P})
d\rho(\bar{t}\to \bar{b}{d}\bar{u})_{\sigma\sigma^\prime}.
\label{eq:AppB_dGammatbar}
\end{align}
\end{subequations}
Note that the decay distributions for $t\to b\bar{s}c$ and $\bar{t}\to\bar{b}s\bar{c}$ are the same as\,(\ref{eq:AppB_dGammat}) and\,(\ref{eq:AppB_dGammatbar}), respectively, where instead of $\bar{d}$ and $d$ momenta we have $\bar{s}$ and $s$ momenta, while the identification probability $P_{\bar{s}c}=P_{s\bar{c}}$ may be significantly larger than 0.5, the most pessimistic value which was assumed in Ref.\,\cite{Hagiwara:2017ban}.

Finally, we find it encouraging that the $t$ and $\bar{t}$ decay angular asymmetries have the same sign when 
\begin{eqnarray}
\vec{P}(\bar{t})\simeq-\vec{P}(t)
\end{eqnarray}
as suggested from approximate CP${\rm \tilde{T}}$ invariance in section\,\ref{section:Todd} and from Figs.\,\ref{fig:pol4} and\,\ref{fig:P2_W} in section\,\ref{sec:pol}. This tells that the polarization asymmetry can be measured even if we cannot distinguish $t$ from $\bar{t}$, which may often be the case for hadronic decays. 

\begin{acknowledgements}
 We are grateful to Tie-Jiun Hou, Junichi Kanzaki and Kentarou Mawatari for helpful discussions.  YZ would like to thank Haider Alhazmi, Sally Dawson, Samuel Lane, Ian Lewis and Kun Liu for valuable suggestions.  KH and YZ thank Tao Han and the members of PITT-PACC and Fermilab Theory Division for their warm hospitality, where part of the present work was carried out. This work has been supported in part by the U.S.\,Department of Energy under contract number DE-SC-0017647. YZ is supported by  the U.S. Department of Energy under grant No. DE-SC0019474.
\end{acknowledgements}


\begin{thebibliography}{9}

\bibitem{Sirunyan:2018hoz} 
  A.~M.~Sirunyan {\it et al.} [CMS Collaboration],
  Phys.\ Rev.\ Lett.\  {\bf 120}, no. 23, 231801 (2018)
  [arXiv:1804.02610 [hep-ex]].
  
\bibitem{Aaboud:2018urx} 
  M.~Aaboud {\it et al.} [ATLAS Collaboration],
  Phys.\ Lett.\ B {\bf 784}, 173 (2018)
  [arXiv:1806.00425 [hep-ex]].
  
  
  
\bibitem{Stirling:1992fx} 
  W.~J.~Stirling and D.~J.~Summers,
  Phys.\ Lett.\ B {\bf 283}, 411 (1992).
  
\bibitem{Bordes:1992jy} 
  G.~Bordes and B.~van Eijk,
  Phys.\ Lett.\ B {\bf 299}, 315 (1993).
 
\bibitem{Aaboud:2018jqu} 
  M.~Aaboud {\it et al.} [ATLAS Collaboration],
  Phys.\ Lett.\ B {\bf 789}, 508 (2019)
  [arXiv:1808.09054 [hep-ex]].
  
\bibitem{Aad:2019lpq} 
  G.~Aad {\it et al.} [ATLAS Collaboration],
  Phys.\ Lett.\ B {\bf 798}, 134949 (2019)
  [arXiv:1903.10052 [hep-ex]].
  
\bibitem{Sirunyan:2018egh} 
  A.~M.~Sirunyan {\it et al.} [CMS Collaboration],
  Phys.\ Lett.\ B {\bf 791}, 96 (2019)
  [arXiv:1806.05246 [hep-ex]].
  

\bibitem{Alwall:2014hca} 
  J.~Alwall {\it et al.},
  JHEP {\bf 1407}, 079 (2014).
 [arXiv:1405.0301 [hep-ph]].
  
\bibitem{Alloul:2013bka} 
  A.~Alloul, N.~D.~Christensen, C.~Degrande, C.~Duhr and B.~Fuks,
  Comput.\ Phys.\ Commun.\  {\bf 185}, 2250 (2014)
  [arXiv:1310.1921 [hep-ph]].
   
\bibitem{Demartin:2015uha}
  F.~Demartin, F.~Maltoni, K.~Mawatari and M.~Zaro,
  Eur.\ Phys.\ J.\ C {\bf 75} (2015) no.6,  267.
  [arXiv:1504.00611 [hep-ph]].
   
\bibitem{Beenakker:2002nc} 
  W.~Beenakker, S.~Dittmaier, M.~Kramer, B.~Plumper, M.~Spira and P.~M.~Zerwas,
  Nucl.\ Phys.\ B {\bf 653}, 151 (2003)
  [hep-ph/0211352].


\bibitem{Tanabashi:2018oca} 
  M.~Tanabashi {\it et al.} [Particle Data Group],
  Phys.\ Rev.\ D {\bf 98}, no. 3, 030001 (2018).
  
\bibitem{Khachatryan:2015ota} 
  {
  V.~Khachatryan {\it et al.} [CMS Collaboration],
  JHEP {\bf 1606}, 177 (2016).
  [arXiv:1509.08159 [hep-ex]].
  }
\bibitem{CMS PAS HIG-17-005} 
 {
[CMS Collaboration],
  CMS PAS HIG-17-005.
  }
  
      \bibitem{CMS PAS HIG-17-009} 
     {
[CMS Collaboration],
  CMS PAS HIG-17-009.
  }
  
  \bibitem{CMS PAS HIG-17-016} 
  {
[CMS Collaboration],
  CMS PAS HIG-17-016.
 }

  
\bibitem{Sirunyan:2018lzm} 
  A.~M.~Sirunyan {\it et al.} [CMS Collaboration],
  Phys.\ Rev.\ D {\bf 99}, no. 9, 092005 (2019)
  [arXiv:1811.09696 [hep-ex]].
  
\bibitem{Maltoni:2001hu} 
  F.~Maltoni, K.~Paul, T.~Stelzer and S.~Willenbrock,
  Phys.\ Rev.\ D {\bf 64}, 094023 (2001).
  [hep-ph/0106293].
  
\bibitem{Barger:2009ky} 
  V.~Barger, M.~McCaskey and G.~Shaughnessy,
  Phys.\ Rev.\ D {\bf 81}, 034020 (2010).
  [arXiv:0911.1556 [hep-ph]].
  
\bibitem{Chang:2014rfa} 
  J.~Chang, K.~Cheung, J.~S.~Lee and C.~T.~Lu,
  JHEP {\bf 1405}, 062 (2014)
  [arXiv:1403.2053 [hep-ph]].
 

\bibitem{Biswas:2012bd} 
  S.~Biswas, E.~Gabrielli and B.~Mele,
  JHEP {\bf 1301}, 088 (2013).
  [arXiv:1211.0499 [hep-ph]].
      
\bibitem{Yue:2014tya} 
  J.~Yue,
  Phys.\ Lett.\ B {\bf 744}, 131 (2015).
  [arXiv:1410.2701 [hep-ph]].
  
\bibitem{Gritsan:2016hjl} 
  A.~V.~Gritsan, R.~Röntsch, M.~Schulze and M.~Xiao,
  Phys.\ Rev.\ D {\bf 94}, no. 5, 055023 (2016).
  [arXiv:1606.03107 [hep-ph]].
   
\bibitem{Farina:2012xp}
  M.~Farina, C.~Grojean, F.~Maltoni, E.~Salvioni and A.~Thamm,
  JHEP {\bf 1305} (2013) 022.
  [arXiv:1211.3736 [hep-ph]].
  
\bibitem{Agrawal:2012ga} 
  P.~Agrawal, S.~Mitra and A.~Shivaji,
  JHEP {\bf 1312}, 077 (2013).
  [arXiv:1211.4362 [hep-ph]].
  
\bibitem{Kobakhidze:2014gqa} 
  A.~Kobakhidze, L.~Wu and J.~Yue,
  JHEP {\bf 1410}, 100 (2014)
  [arXiv:1406.1961 [hep-ph]].

\bibitem{Atwood:2000tu} 
  D.~Atwood, S.~Bar-Shalom, G.~Eilam and A.~Soni,
  Phys.\ Rept.\  {\bf 347}, 1 (2001)
  [hep-ph/0006032].
  
\bibitem{Rindani:2016scj} 
  S.~D.~Rindani, P.~Sharma and A.~Shivaji,
  Phys.\ Lett.\ B {\bf 761}, 25 (2016).
  [arXiv:1605.03806 [hep-ph]].
  

\bibitem{Kraus:2019myc} 
  M.~Kraus, T.~Martini, S.~Peitzsch and P.~Uwer,
  arXiv:1908.09100 [hep-ph].
  
  
\bibitem{Djouadi:2005gj} 
 {A.~Djouadi,
  Phys.\ Rept.\  {\bf 459}, 1 (2008).}
  [hep-ph/0503173].
  
\bibitem{Barger:2018tqn} 
  V.~Barger, K.~Hagiwara and Y.~J.~Zheng,
  Phys.\ Rev.\ D {\bf 99}, no. 3, 031701 (2019).
  [arXiv:1807.00281 [hep-ph]].
  
\bibitem{Faroughy:2019ird} 
  D.~A.~Faroughy, J.~F.~Kamenik, N.~Ko{\v{s}}nik and A.~Smolkovi{\v {c}},
  arXiv:1909.00007 [hep-ph].
  
\bibitem{Hagiwara:2009wt} 
  K.~Hagiwara, Q.~Li and K.~Mawatari,
  JHEP {\bf 0907}, 101 (2009).
  [arXiv:0905.4314 [hep-ph]].
    
\bibitem{Hagiwara:1985yu} 
  K.~Hagiwara and D.~Zeppenfeld,
  Nucl.\ Phys.\ B {\bf 274}, 1 (1986).
  
        
      \bibitem{Murayama:1992gi} 
  K.~Hagiwara, H.~Murayama and I.~Watanabe,
  Nucl.\ Phys.\ B {\bf 367}, 257 (1991).
  H.~Murayama, I.~Watanabe and K.~Hagiwara,
  KEK-91-11.
 
  
\bibitem{Hagiwara:2017ban}
  K.~Hagiwara, H.~Yokoya and Y.~J.~Zheng,
  JHEP {\bf 1802} (2018) 180.
  [arXiv:1712.09953 [hep-ph]].
 
\bibitem{Dulat:2015mca} 
  S.~Dulat {\it et al.},
  Phys.\ Rev.\ D {\bf 93}, no. 3, 033006 (2016).
  [arXiv:1506.07443 [hep-ph]].
   
\bibitem{Atwood:1992vj} 
  D.~Atwood, A.~Aeppli and A.~Soni,
  Phys.\ Rev.\ Lett.\  {\bf 69}, 2754 (1992).
  
\bibitem{DeRujula:1978bz} 
  A.~De Rujula, R.~Petronzio and B.~E.~Lautrup,
  Nucl.\ Phys.\ B {\bf 146}, 50 (1978).
  
\bibitem{Hagiwara:1982cq} 
  K.~Hagiwara, K.~i.~Hikasa and N.~Kai,
  Phys.\ Rev.\ D {\bf 27}, 84 (1983).

\bibitem{Hagiwara:1986vm} 
  K.~Hagiwara, R.~D.~Peccei, D.~Zeppenfeld and K.~Hikasa,
  Nucl.\ Phys.\ B {\bf 282}, 253 (1987).
  
\bibitem{Barger:2020} 
  V.~Barger, K.~Hagiwara and Y.~J.~Zheng, in preparation.
 
 \end{thebibliography}
\end{document}